\documentclass[aps,prapplied,twocolumn,groupedaddress]{revtex4-2}     %
\usepackage[utf8]{inputenc}			%
\usepackage{graphicx}				%
\usepackage{hyperref}
\hypersetup{colorlinks=true, linkcolor=blue, urlcolor=blue, citecolor=blue, final}
\usepackage{tabularx}
\usepackage{multirow}
\usepackage{array}
\usepackage[percent]{overpic}
\usepackage[caption=false]{subfig}
\usepackage{asymptote}
\usepackage{rotating}
\usepackage{epsfig}			%
\usepackage{xcolor}			%
\usepackage{amsmath}			%
\usepackage{amssymb}			%
\usepackage{bm}				%
\usepackage{units}			%
\usepackage{enumitem}			%
\usepackage{pict2e}

\newcolumntype{L}{>{$}l<{$}}
\newcolumntype{C}{>{$}c<{$}}
\newcolumntype{R}{>{$}r<{$}}

\definecolor{darkgreen}{RGB}{0,175,0}
\definecolor{navy}{RGB}{0,0,150}
\definecolor{deepgreen}{RGB}{0,100,0}
\definecolor{heavycyan}{cmyk}{255,0,0,.175}

\begin{document}

\newcommand{\Ibd}{Inverse beta decay (IBD)\renewcommand{\ibd}{IBD}}
\newcommand{\ibd}{inverse beta decay (IBD)\renewcommand{\ibd}{IBD}}
\newcommand{\mtc}{miniTimeCube (mTC)\renewcommand{\mtc}{mTC}}
\newcommand{\uh}{University of Hawai'i (UH)\renewcommand{\uh}{UH}}
\newcommand{\llnl}{Lawrence Livermore National Laboratory (LLNL)\renewcommand{\llnl}{LLNL}}
\newcommand{\dnn}{Defense Nuclear Nonproliferation (DNN)\renewcommand{\dnn}{DNN}}
\newcommand{\nnsa}{National Nuclear Security Administration (NNSA)\renewcommand{\nnsa}{NNSA}}
\newcommand{\pmt}{photomultiplier tube (PMT)\renewcommand{\pmt}{PMT}}
\newcommand{\pmts}{photomultiplier tubes (PMTs)\renewcommand{\pmts}{PMTs}}
\newcommand{\sipm}{silicon photomultiplier (SiPM)\renewcommand{\sipm}{SiPM}}
\newcommand{\sipms}{silicon photomultipliers (SiPMs)\renewcommand{\sipms}{SiPMs}}
\newcommand{\rat}{Reactor Analysis Tool Plus Additional Code (RAT-PAC)\renewcommand{\rat}{RAT-PAC}}
\newcommand{\mc}{Monte Carlo (MC)\renewcommand{\mc}{MC}}
\newcommand{\dc}{Double Chooz}
\newcommand{\santa}{Segmented AntiNeutrino Tomography Apparatus (SANTA)\renewcommand{\santa}{SANTA}}
\newcommand{\nl}{Neutrino Lattice (NuLat)\renewcommand{\nl}{NuLat}}
\newcommand{\songs}{San Onofre Nuclear Generating Station (SONGS)\renewcommand{\songs}{SONGS}}
\newcommand{\sandd}{Segmented Anti-Neutrino Directional Detector (SANDD)\renewcommand{\sandd}{SANDD}}
\newcommand{\rol}{Raghavan Optical Lattice (ROL)\renewcommand{\rol}{ROL}}
\newcommand{\raa}{Reactor Antineutrino Anomaly (RAA)\renewcommand{\raa}{RAA}}
\newcommand{\nist}{National Institute for Standards and Technology (NIST)\renewcommand{\nist}{NIST}}
\newcommand{\ncnr}{National Center for Neutron Research (NCNR)\renewcommand{\ncnr}{NCNR}}
\newcommand{\mcps}{Micro-channel-plate Photomultiplier Tubes (MCP-PMTs)\renewcommand{\mcps}{MCP-PMTs}}
\newcommand{\mcp}{Micro-channel-plate Photomultiplier Tube (MCP-PMT)\renewcommand{\mcp}{MCP-PMT}}
\newcommand{\wbls}{water-based liquid scintillator (WBLS)\renewcommand{\wbls}{WBLS}}
\newcommand{\ols}{organic liquid scintillator (OLS)\renewcommand{\ols}{OLS}}
\newcommand{\psd}{pulse-shape discrimination (PSD)\renewcommand{\psd}{PSD}}
\newcommand{\tir}{total internal reflection (TIR)\renewcommand{\tir}{TIR}}
\newcommand{\fov}{field-of-view (FOV)\renewcommand{\fov}{FOV}}

\newcommand{\nb}{$\overline{\nu_e}$}				%
\newcommand{\nn}{$n$}						%
\newcommand{\ep}{$e^+$}						%
\newcommand{\nbm}{\overline{\nu_e}}				%
\newcommand{\nnm}{n}						%
\newcommand{\epm}{e^+}						%
\newcommand{\recnc}{$\vec{S_R}$}        %
\newcommand{\rhatnc}{$\hat{S_R}$}       %
\newcommand{\snunc}{$\hat{S_T}$}        %
\newcommand{\recmnc}{\vec{S_R}}         %
\newcommand{\rhatmnc}{\hat{S_R}}        %
\newcommand{\snumnc}{\hat{S_T}}         %
\newcommand{\nbc}{\nb}
\newcommand{\nnc}{\nn}
\newcommand{\epc}{\ep}
\newcommand{\nbcm}{\nbm}
\newcommand{\nncm}{\nnm}
\newcommand{\epcm}{\epm}
\newcommand{\rec}{\recnc}
\newcommand{\rhat}{\rhatnc}
\newcommand{\snu}{\snunc}
\newcommand{\recm}{\recmnc}
\newcommand{\rhatm}{\rhatmnc}
\newcommand{\snum}{\snumnc}

\newcommand{\dash}{\textbf{---}\ }				%
\newcommand{\bs}{\boldsymbol}					%
\newcommand{\nucl}[2]{{}^{#1}\mbox{#2}}				%

\newcommand{\titletext}{Directional Response of Several Geometries for Reactor-Neutrino Detectors}
\title{\titletext}                %

\author{Mark~J.~Duvall}
\email{mjduvall@hawaii.edu}
\author{Brian~C.~Crow}
\author{Max~A.A.~Dornfest}
\author{John~G.~Learned}
\affiliation{Department of Physics and Astronomy, University of Hawai'i at M\={a}noa}
\author{Marc~F.~Bergevin}
\author{Steven~A.~Dazeley}
\author{Viacheslav~A.~Li}
\affiliation{Lawrence Livermore National Laboratory}

\date{\today}

\begin{abstract}
We report simulation studies of six low-energy electron-antineutrino detector designs, with the goal of determining their ability to resolve the direction to an antineutrino source.
Such detectors with target masses on the one-ton scale are well-suited to reactor monitoring at distances of 5--25 meters from the core.
They can provide accurate measurements of reactor operating power, fuel mix, and burnup, as well as unsurpassed nuclear non-proliferation information in a non-contact cooperating reactor scenario such as those used by IAEA.
A number of groups around the world are working on programs to develop detectors similar to some of those in this study.

Here, we examine and compare several approaches to detector geometry for their ability not only to detect the inverse beta decay (IBD) reaction, but also to determine the source direction of incident antineutrinos.
The information from these detectors provides insight into reactor power and burning profile, which is especially useful in constraining the clandestine production of weapons material.
In a live deployment, a non-proliferation detector must be able to isolate the subject reactor, possibly from a field of much-larger power reactors; directional sensitivity can help greatly with this task.
We also discuss implications for using such detectors in longer-distance observation of reactors, from a few km to hundreds of km.

We have modeled six abstracted detector designs, including two for which we have operational data for validating our computer modeling and analytical processes.
We have found that the most promising options, regardless of scale and range, have angular resolutions on the order of a few degrees, which is better than any yet achieved in practice by a factor of at least two.
\end{abstract}

\maketitle %
\tableofcontents

\section{\label{sec:intro}Introduction}

The first detection of neutrinos was accomplished by Reines and Cowan in the mid-1950s using a nuclear reactor as a source~\cite{Cowan:1992xc}.
Rapidly thereafter, neutrinos were observed from cosmic rays (South Africa~\cite{Reines:1965qk} and Kolar Gold Mines~\cite{Achar1965}) and at accelerators (Brookhaven~\cite{PhysRevLett.15.42}); many observations were successfully performed in the proximity of reactors around the world; and solar neutrinos were observed in the Homestake mine~\cite{PhysRevLett.20.1205}.
Through these experiments and others, weak interaction theory was validated. Limits were set on neutrino mass, neutrino stability, and the number of neutrino types. %

Soon thereafter, in the 1960s and the 1970s, commercial power reactors began producing power in quantities enough to start considering neutrinos as a utility~\cite{Borovoi1978} and a neutrino-oscillation study tool. In the 1980s and early 1990s, there were a number of experiments at reactors carried out at multiple baselines to look for neutrino-oscillations effect (Rovno~\cite{Afonin:1987gi}, G\"osgen~\cite{Gabathuler:1984gs}, Bugey~\cite{Cavaignac:1984sp}, Krasnoyarsk~\cite{Vidyakin:1994ut}, and Savannah River~\cite{Greenwood:1996pb}).
In 1998, SuperKamiokande detected the first clear example of neutrino oscillations via the disappearance of cosmic-ray muon neutrinos traversing the Earth~\cite{PhysRevLett.81.1562}, setting off the modern avalanche of neutrino studies.
Next, the SNO experiment found evidence of both solar electron-neutrino disappearance and total solar-neutrino conservation via observation of the unsuppressed neutral-current solar-neutrino interactions \cite{PhysRevLett.87.071301,Aharmim:2013}.
In 2003, the KamLAND experiment detected the disappearance of neutrinos from the ensemble of reactors around Japan, producing the first unequivocal evidence that electron neutrinos oscillate~\cite{PhysRevLett.90.021802}. 
This cemented the case for oscillatory neutrino flavor changing, explaining the decades-long solar-neutrino quandary.
The subject of using antineutrinos in the nonproliferation context has seen a renaissance in the recent years~\cite{PhysRevLett.113.042503,Bernstein:2019hix}. The near-field and mid-field the community has entered a precision era as seen with PROSPECT \cite{PROSPECT:2020sxr}  and DayaBay ~\cite{PhysRevLett.131.021802, PhysRevLett.130.161802}. While the long-range detection is becoming possible --- with the undoped liquid scintillator KamLAND~\cite{PhysRevLett.90.021802} and water SNO+ detector~\cite{PhysRevLett.130.091801}, with two other detectors, JUNO and the gadolinium-doped SuperKamiokande detector, coming online.%

For a time, the concept of long-range ($\gtrsim\mathrm{10~km}$) reactor monitoring was dismissed as impossible due to the large (kiloton-scale) antineutrino detectors required. 
The successful operation of large nucleon-decay search instruments in the 1980s (IMB, Kamiokande, Baksan, Kolar, Frejus) \cite{BERGER1989489}.%
The concomitant demonstrated sensitivity to MeV-scale antineutrinos, shown particularly well by observations of \nb\ from Supernova 1987A, provided strong evidence for this possibility~\cite{PhysRevLett.58.1494,PhysRevLett.58.1490}. %
The first report examining long-range detection of reactors was presented in 2003, in a study of hypothetical possibilities for long-range monitoring of reactors, without the consideration of size and cost, simply to examine extreme limits on monitoring~\cite{PhysRevLett.90.021802}.
The conclusion was that given extrapolations of technology to perhaps 50 years out, and development of several large underwater IBD detectors in the megaton class, it would be feasible to monitor reactors anywhere on Earth except in some central continental regions.

A substantial study published in \textit{Physics Reports} in 2015~\cite{Jocher:2013gta} demonstrated that a relatively small reactor could be distinguished from a reactor complex with two or three observing stations (hundred-kiloton scale) at substantial distances (hundreds of km).
But we also learned that a small 50-$\mathrm{MW_{Th}}$ weapons-grade fuel-producing facility of only {1\%} or so of the total reactors within detection range would be nearly impossible to discriminate individually. %
However, this paper also observed that even modest antineutrino-direction resolution would greatly improve the ability to discern and monitor a clandestine reactor.
The study that is reported herein follows this lead, starting with smaller detectors scaled around one ton or less.

A key difficulty in such studies is the relatively low amount of energy deposited following an \ibd\ interaction. 
With energy deposition typically ranging around only a few MeV per event, the reactor signal can be completely swamped by backgrounds.
Almost all reactors are built near the Earth's surface (and near cooling / water resources) for obvious practical reasons.
Hence, the optimal local-monitoring detector would be located directly under the reactor it is observing, but such an arrangement is rarely practical.
For longer-range monitoring, however, cavern- or ocean-located detectors can eliminate the backgrounds generated by cosmic-ray muons, which are dominant in near-surface detectors.
There is a limit even to this, however, at depths of $\gtrsim4000~\mathrm{m.w.e.}$, where local nuclear-decay-generated backgrounds come to dominate as detector size scales upwards~\cite{PhysRevLett.130.091801}. 
But the corresponding range limit may be several tens~of~kilometers or hundreds of kilometers depending on the power of the reactor, backgrounds, and the size and type of the detector~\cite{Li_Gd_water, PhysRevLett.90.021802}.
Moreover, other recent work on the total flux of antineutrinos from all the world's reactors will present a limit on the observation range of even the largest and deepest individual conceivable reactors, probably in the hundred-km range~\cite{Barna_2015}.
We should mention that along with directionality, the observation of the antineutrino spectrum leads to the possibility of blind antineutrino-source range determination (without foreknowledge of the source or distance)~\cite{Smith:2010zzb}~\cite{Snowmass:2021}.
Given statistics of several thousand events, the Fourier transform of the inverse antineutrino spectrum (on distance divided by energy) yields a unique measure of the source-detector distance~\cite{Wilson:2023owp}.
To make this practical, one needs both excellent IBD energy resolution \dash\ in the range of a few percent \dash\ and a low energy threshold, going well below the spectrum peak around 4 MeV, usually background-limited.
We do not focus on these matters herein but concentrate on the directionality issue.
Ultimately, budget issues will limit practical plans; the strategy here is to concentrate on achievable performance and then later to address these other matters.
The full-scale realization of these futuristic detectors may eventually require technological advancements in target materials, optical detectors, electronics, and even deep-ocean IBD vessels.
However, we show in this study that novel approaches to detector geometry should make substantial performance improvements achievable using only currently-existing technology.

There are also other applications for such detectors, from geoneutrino studies~\cite{deMeijer:2005zz,SHIMIZU2007147,Watanabe:2014} to directionally-informed early warning for supernovae~\cite{Beacom:1998fj} and galactic-distance gravitational events, for example.
It should be noted that any of these reactor-monitoring detectors can be actively utilized for important physics and astrophysics purposes simultaneously with the monitoring mission at little or probably no mission compromise or extra expense.
The antineutrino rates are so low that no sensitivity is lost in monitoring for supernovae and other phenomena.
A unique ability could come from early warning of not only an increasing rate of IBD events but the unique ability to give the first clue as to the direction of the incipient supernova.
No existing or proposed detector can claim this early-SN-direction capability.

\section{\label{sec:dirn}Directionality from Inverse Beta Decay}

\subsection{\label{subsec:dirn_def}Definition}
We start with defining the coordinate system used herein.
Under the assumption that the reactors in question are located in the same horizontal plane as the detector (neither in the sky nor deep underground relative to the observation point), we focus on determination of the azimuthal direction.

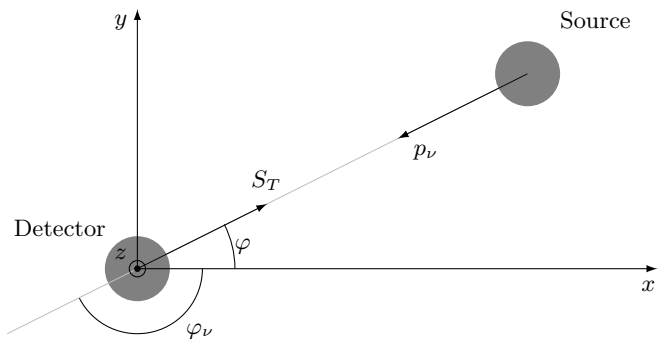
\begin{figure}[ht]
\setlength{\unitlength}{.1\linewidth}
\begin{picture}(10,5)
  \put(2,1){\color{gray}\circle*{1}} %
  \put(8,4){\color{gray}\circle*{1}} %
  \put(0,0){\color{lightgray}\line(8,4){8}}
  \put(2,1){\vector(1,0){8}}  %
  \put(2,1){\vector(0,1){4}}  %
  \put(8,4){\vector(-8,-4){2}} %
  \put(6.25,2.75){\text{$p_\nu$}}
  \put(2,1){\vector(8,4){2}} %
  \put(3.75,2.25){\text{$S_T$}}
  \put(2,1){\circle{0.25}}
  \put(2,1){\color{black}\circle*{0.1}}
  \put(9.75,0.65){\texttt{$x$}}
  \put(1.65,4.75){\texttt{$y$}}
  \put(1.65,1.15){\texttt{$z$}}
  \put(2,1){\arc[0,27]{1.5}} %
  \put(3.5,1.3){\texttt{$\varphi$}}
  \put(2,1){\arc[0,-153]{1.}} %
  \put(2.75,0){\texttt{$\varphi_\nu$}}
  \put(0.1,1.5){\color{black}\text{Detector}}
  \put(8.5,4.7){\color{black}\text{Source}}
\end{picture}
        \caption{Diagram of the key directional quantities ($xy$-projection). The vector $S_T$ is a vector indicating the true direction towards the source. In this study, all antineutrinos are assumed to have the same direction in their momentum $p_\nu$.}
\label{fig:phi_def}
\end{figure}

\begin{gather}
    \varphi_{\nu}\ =\ \arctan \left(\frac{p_{\nu\perp}}{p_{\nu\parallel}} \right)\ =\ \arctan \left( \frac{\Delta y}{\Delta x} \right) \label{eq:phi_calc} ,\\
    \varphi\ \equiv\ 180^\circ - \varphi_{\nu}   \label{eq:phi_def}
\end{gather}

In detector physics, \emph{directionality} refers to the ability of a detector to determine the incident direction of an incoming particle, and by extension to infer the direction pointing from the detector to the particle's source.
Fig.~\ref{fig:phi_def} illustrates this concept as applied to a detector attempting to reconstruct the direction \snu\ to the source of the incoming antineutrinos.
As mentioned above, the source is assumed to be located on the horizon relative to the detector (i.e., $z_{Detector}\ =\ z_{Source}$), so this problem reduces to reconstructing the azimuthal angle $\varphi$, which can be found as given in Eqs.~\ref{eq:phi_calc} and~\ref{eq:phi_def}.
Note that some of the background from distant reactors will be substantially out-of-plane, so we get some small relief from backgrounds which are due to reactors located well below the horizon. 
This effect is mostly relevant for large, long-baseline detectors\cite{DanielsonAAP2018}.

\subsection{\label{subsec:dirn_meth}Acquisition}

\ibd\ is a charged-current weak interaction in which an electron-antineutrino interacts with a proton by exchanging a $W^+$ boson, producing a positron and a neutron, as follows:
\begin{equation}
    \nbm\ +\ p ~\to~ \epm +\ \nnm
    \label{eq:ibd}
\end{equation}

\begin{figure}[ht]
  \begin{center}
  \includegraphics[width=1.0\linewidth]{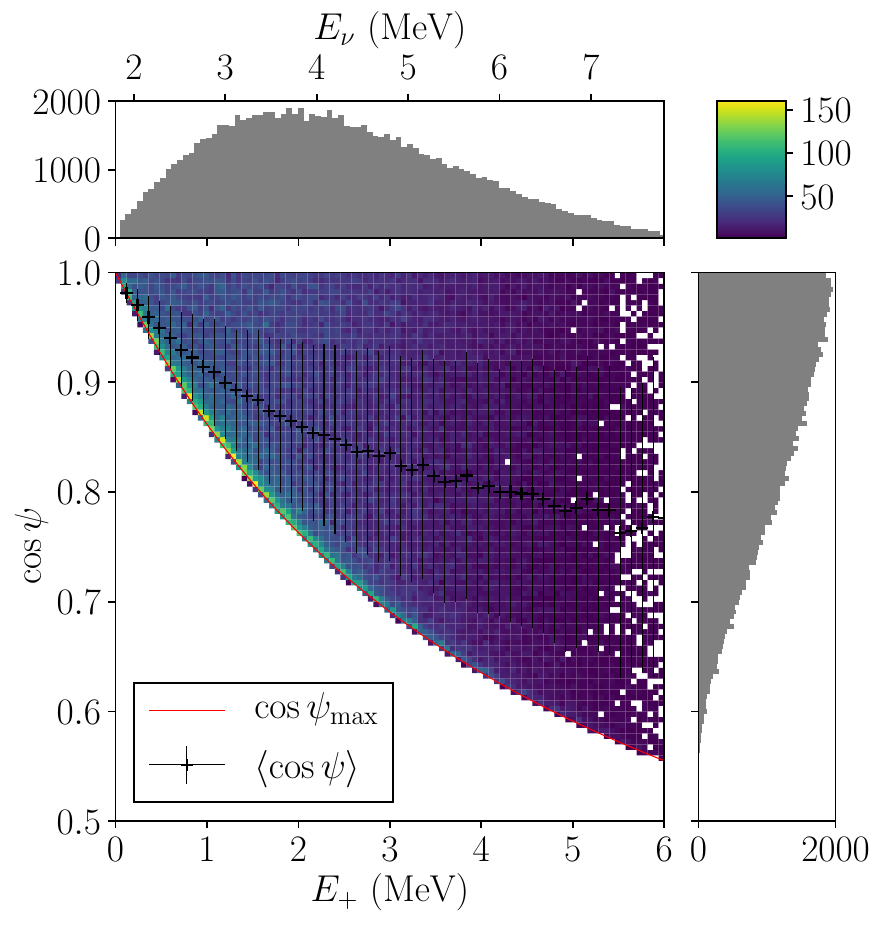}
  \caption{Distribution of positron kinetic energy $E_+$ and cosine of neutron scattering angle $\psi$ relative to the incoming antineutrino, based on simulating 100,000 IBD positron-neutron pairs. The binning in both axes, $\cos \psi$ and positron energy $E_+$, is 100; the projections are shown in gray in the inset 1D histograms. The mean values and standard deviations, along with the analytical calculation for the maximum angle, are also displayed. It is worth noting that the 2D distribution has a sharp peak when the angle approaches the maximum allowed angle (shown in red curve), resulting in the  shift of the mean value towards that curve.}%
  \label{fig_cosPsi_vs_eKE}
  \end{center}
\end{figure}

The kinematics of \ibd, which superficially resemble that of a ping-pong ball (lepton) scattering off of a bowling ball (hadron), have two primary consequences for our detection scheme:
\begin{enumerate}
  \item The \epc\ receives most of the \nbc's kinetic energy (minus 1.8~MeV), and
  \item The \nnc\ receives most of the \nbc's momentum and therefore its directional information.
\end{enumerate}
These can be seen in Fig.~\ref{fig_cosPsi_vs_eKE}.
The key task for \ibd\ directionality is therefore to find the direction of the neutron's momentum as it exits the \ibd\ vertex.
In practice, this is achieved by finding the average direction of neutron displacement from production to capture over a set of events.

\begin{figure}[ht]
\setlength{\unitlength}{.1\linewidth}
\begin{picture}(10,4)
  \put(10,3){\color{gray}\line(-1,0){4}} %
  \put(10,3){\vector(-1,0){1.5}} %

  \put(1,3){\vector(1,0){1}} %
  \put(1,3){\vector(0,1){1}} %
  \put(1,3){\vector(-1,-2){.3}} %
  \put(1.75, 2.65){$x$}
  \put(0.65, 3.65){$y$}
  \put(0.45, 2.45){$z$}

  \put(6,3){\color{black}\circle*{.25}} %
  \put(6.5,1.5){\color{black}\circle*{.25}} %
  \put(1,.5){\color{black}\circle*{.25}} %

  \put(6,3){\vector(-1,-0.2){1}} %
  \put(6,3){\vector(.5,-1.5){.25}} %
  \put(6,3){\line(.5,-1.5){.5}} %
  \put(5.7, 3.5){\text{IBD}}
  \put(5., 3.){\text{n}}
  \put(8.6, 3.5){$\bar\nu_e$}
  \put(6.4,2.4){\text{e}$^+$}
  \put(7.,1.3){\text{e}$^+$ \text{annihilation}}
  \put(0,0){\text{neutron capture}}

  \put(1,.5){\vector(1,0){4}} %
  \put(1,.5){\color{gray}\line(6.5,1.9){5.3}} %
  \put(1,.5){\vector(6.5,1.9){3.5}} %
  \put(4.7,0){$S_T$}
  \put(3.7,1.8){$S_R$}

  \put(2.7,0.7){\texttt{$\psi$}}
  \put(1,.5){\arc[0,16.5]{1.55}} %
  \put(1,.5){\arc[0,16.5]{1.5}} %
\end{picture}
        \caption{Diagram of how the angle $\psi$ between reconstructed $S_R$ and true $S_T$ source direction is defined.
        The vector $S_R$ points from the reconstructed vertex of the neutron capture towards the reconstructed vertex where positron deposited its energy before annihilating. The MeV-scale positron travels a few millimeters before being annihilated with an electron.
        The keV-scale neutron loses its energy in the detector medium before being captured; the separation between IBD vertex and the capture location could be made large if there is no medium in between, as in some of the detector designs considered in this study.
        The antineutrino source was positioned along the $x$-axis: angle $\psi = 0$ would indicate a perfect reconstruction of direction.}
\label{fig:recon}
\end{figure}
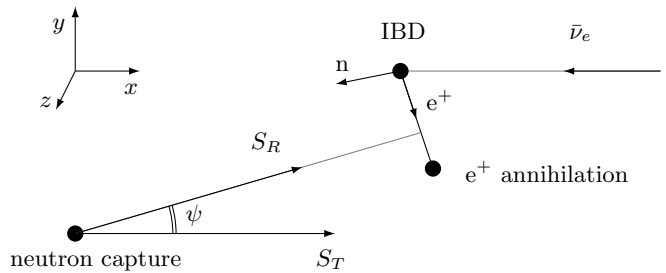

The \epc's displacement from its production (which is at the \ibd\ vertex) to its annihilation is typically much shorter than the \nnc's capture displacement, so we can reliably use a point along the \epc\ track to approximate the location of the \ibd\ vertex.
In other words, the short distance between the \ibd\ vertex and the end of the \epc\ track is negligible compared to the neutron-capture distance.
It is, in fact, so short that it is too small to resolve in practice.  %
The method for extracting directional information from an \ibd\ detector is then to draw a vector from the location of neutron capture to the location of the positron event.
This vector, which we will notate as \rec, represents the reconstructed direction to the antineutrino source.
The true direction from the detector back to the antineutrino source will be indicated by the unit vector \snu.
This process for a single typical \ibd\ event is shown in Fig.~\ref{fig:recon}:
\begin{enumerate}
  \item An incoming \nbc\ enters an arbitrary region of the detector's target material.
  \item The \nbc\ undergoes \ibd\ by interacting with a $p^+$ in the target material, producing a \nnc\ and an \epc at the expense of the neutron-proton mass difference of 1.8 MeV in incoming antineutrino energy.
  \item The \epc\ travels a few mm, leaving a trail of ionization, and annihilates with an $e^-$ into two 511-keV $\gamma$; this is collectively called the \textit{prompt event}. We locate the prompt event at the energy centroid (midpoint) of the \epc 's ionization track. The $\gamma$ have a much longer interaction length and do not contribute substantially to the prompt energy centroid.
  \item Meanwhile, the \nnc\ undergoes a series of intermediate scattering steps (omitted from Fig.~\ref{fig:recon} for clarity) and eventually captures, either on a free proton (\textit{np}) or usually on a nucleus of dopant material; this is called the \textit{delayed event} and is typically $\mathcal{O}(10~\mu\mathrm{s})$ later in the relevant materials. This is much later than the prompt event, which is delayed by only a few {ns} from the antineutrino interaction.
  \item We draw our vector \rec\ from the delayed event to the prompt event.
  \item Finally, we take as our reconstructed source direction $\rhatm \equiv \recm / \| \recm \|$.
\end{enumerate}
In a directionally-sensitive detector, the average of \rhat\ over a large number of events will increasingly tend toward \snu.
The magnitude $\| \recm \|$ gives us the neutron's approximate capture distance, which we will hereafter denote as $d_n$ (\textit{i.e.,} $d_n \equiv \| \recm \|$).
We will sometimes be interested in this, but for directional purposes it is often convenient to simply use the corresponding unit vector, \rhat.
We can also define the quantity $\psi$ as the angle between \rhat\ and \snu, as shown in Fig.~\ref{fig:recon} and according to the following:

\begin{equation}
  \cos \psi = \hat S_R \cdot \hat S_T
  \label{eq:psi_def}
\end{equation}

In this work, for the 2D detectors --- angle $\psi$ has only the azimuthal component (reducing $S_R$ and $S_T$ to their projections in $xy$-plane); while for 3D detectors --- both azimuthal and zenith.

\subsection{\label{subsec:dirn_stats}Statistics}
The determination of the direction of the incoming antineutrino is an inherently stochastic process.
This is because of the scattering undergone by the neutron during its moderation / thermalization phase. %
This scattering is a \emph{nearly}-isotropic random-walk after the first scatter, with only a slight bias in the forward direction of antineutrino travel.
The neutron's capture displacement (and consequently \rhat) for any particular event will then be in a nearly-random direction, and it is therefore only the average of \rhat\ over a substantial number of events that will reliably point in a direction near \snu\ and give a meaningful result for the direction to the antineutrino source.
This is true even for cases in which the \nnc-scattering region can be separated from the \epc-track region, as will be discussed in detail in \S\ref{subsec:sep}.
The results in~\cite{Vogel:1999zy} also indicate that on average, the \epc\ will have a slight bias to leave the \ibd\ interaction
  heading in the backward direction relative to the incoming \nbc.

\subsection{\label{subsec:mat_cap}Neutron-Capture Agent}
The dopant isotope determines the inter-event time and both the energy and degree of localization for the delayed event, along with a host of other design considerations.
Common dopant isotopes include $\nucl{6}{Li},\ \nucl{10}{B},\ \textrm{and}\ \nucl{155}{Gd}/\nucl{157}{Gd}$, all three of which are used in this study.
Neutron captures on $\nucl{1}{H}$ are also common, simply because frequently-used target materials such as water and polymers contain hydrogen.
The natural isotopic abundances~\cite{Zucker:99} and capture reactions for the above nuclei are given in Eqs.~\ref{eq:cap_li}~\cite{Knoll:2000fj}, \ref{eq:cap_b}~\cite{mTC:2016yys}, and \ref{eq:cap_gd}~\cite{Knoll:2000fj}~\cite{Tanaka:2019}~\cite{Hagiwara:2019}.

$\nucl{1}{H}~(>99.9\%):$
\begin{equation}
  n + \nucl{1}{H} \to \nucl{2}{H} + \gamma\mathrm{'s} (2200\; \mathrm{keV~total}).
  \label{eq:cap_h}
\end{equation}

$\nucl{6}{Li}~(7.6\%):$
\begin{equation}
  n + \nucl{6}{Li} \to \nucl{3}{H} (2730\; \mathrm{keV}) + \nucl{4}{He} (2050\; \mathrm{keV}).
  \label{eq:cap_li}
\end{equation}

$\nucl{10}{B}~(20.0\%):$
\begin{equation}
  \begin{split}
    n + \nucl{10}{B} \to & \nucl{7}{Li} (1015\; \mathrm{keV}) + \nucl{4}{He} (1775\; \mathrm{keV}), \quad \sim 6 \% \\%
      \to & \nucl{7}{Li}^* + \nucl{4}{He} (1471\; \mathrm{keV}), \qquad \qquad \quad \sim 94 \% \\%
      & \hookrightarrow \nucl{7}{Li}^* \to \nucl{7}{Li} (839\; \mathrm{keV}) + \gamma (478\; \mathrm{keV}).
  \end{split}
  \label{eq:cap_b}
\end{equation}

$\nucl{155}{Gd}~(14.8\%)~~\&~~\nucl{157}{Gd}~(15.7\%):$
\begin{subequations}
  \begin{align}
    n + \nucl{155}{Gd} \to \nucl{156}{Gd} + \gamma\mathrm{'s} (8500\; \mathrm{keV~total}) \label{subeq:cap_gd155}, \\
    n + \nucl{157}{Gd} \to \nucl{158}{Gd} + \gamma\mathrm{'s} (7900\; \mathrm{keV~total}) \label{subeq:cap_gd157}.
  \end{align}
  \label{eq:cap_gd}
\end{subequations}

\subsection{\label{subsec:dirn_prev}Other Related Detectors}

\textbf{Previous}
The forward/backward asymmetry in IBD neutrons was first discovered by Gabrielle Zacek by analyzing data from the G\"osgen experiment~\cite{Zacek:1984qi}. %
The effect was also observed by Palo Verde~\cite{PaloVerde_direction} and Bugey 3.
The first experiment to successfully demonstrate directional detection of electron antineutrinos via \ibd\ was the Chooz experiment, which used $\sim 2700$ \ibd\ events to ``[locate] an antineutrino source to within a cone of half-aperture $\approx 22^o$ at the 68\% C.L.'' (confidence level)~\cite{Apollonio:1999jg}.
The \dc\ (DC) experiment then improved on this result, acquiring $\sim$ 8200 \ibd\ events and reducing the cone to half-aperture $\approx 6^o$~\cite{cadenThesis, Caden:2012bm}. %

\textbf{Concurrent and Future}
Other collaborations working along these lines include the PROSPECT~\cite{PROSPECT:2020sxr} and Daya Bay experiments, both have published their final measurements last year~\cite{PhysRevLett.131.021802,PhysRevLett.130.161802}.
The RENO~\cite{RENO:2018pwo} (Korea), Double Chooz~\cite{Abe:2014bwa} (France), and Borexino~\cite{Bellini:2013pxa} experiments have concluded.
KamLAND and SuperKamiokande are operating, but neither have the resolution to resolve IBD directions.
The JUNO project in China is under construction and hopefully will soon operate.

\section{\label{sec:detectors}Detector Geometries}

\subsection{\label{subsec:mono}Monolithic / Single-Volume}

Because of \dc' success as mentioned above, we have chosen it as a basis for comparison in this study.
In Chooz and \dc, as well as other major antineutrino experiments such as RENO~\cite{RENO:2018pwo} and Daya Bay~\cite{DayaBay:2012aa}, all of the target material is arranged in a single, central volume. We will hereafter refer to this basic design as the \textbf{monolithic} type.
This volume is usually a simple geometric shape such as a sphere, cube, or cylinder. Choosing such highly-symmetric configurations simplifies reconstruction of the interaction locations.

\begin{figure}[ht]
  \begin{center}
  \includegraphics[width=1.0\linewidth]{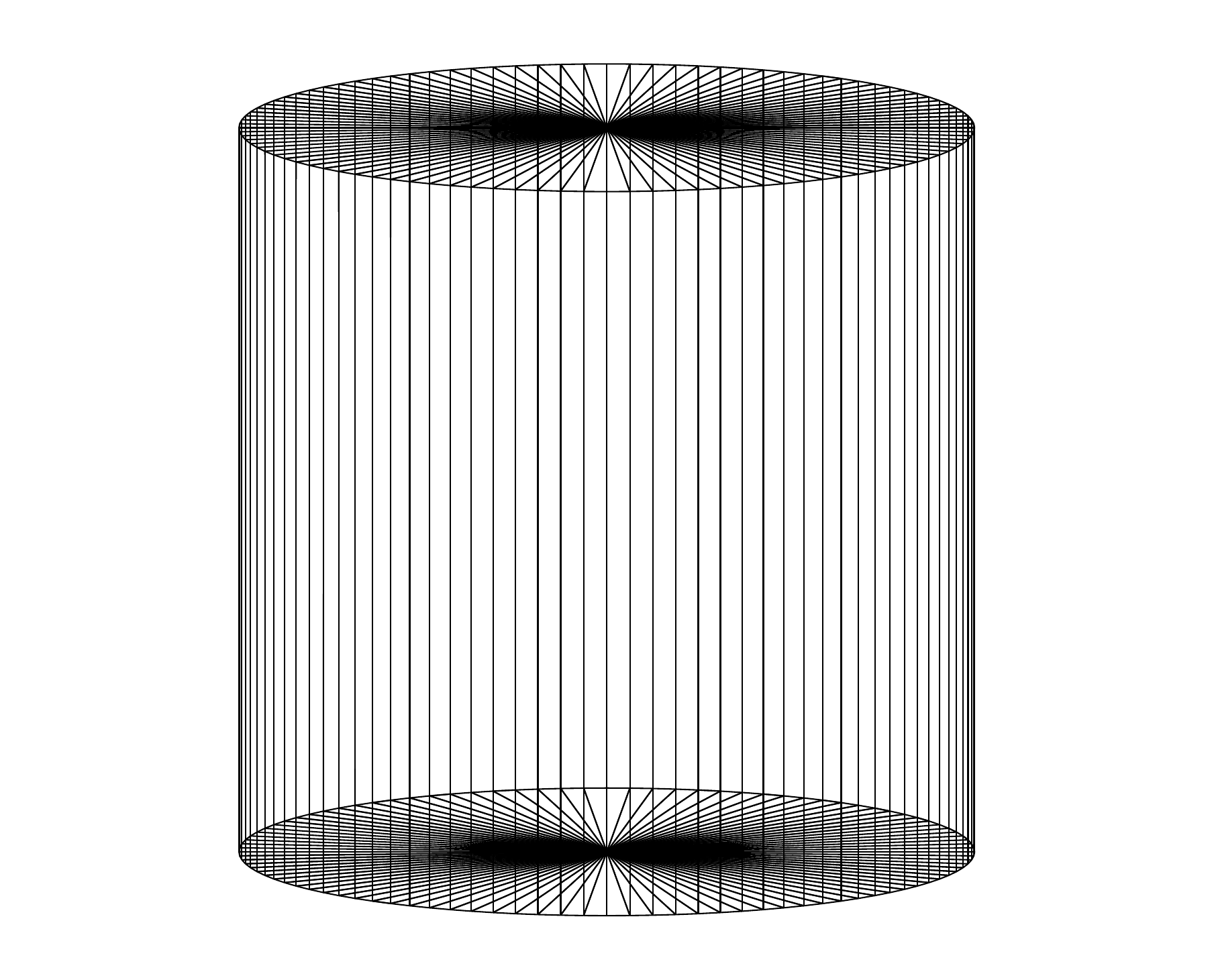}
  \caption{Simulated Monolithic Geometry --- a right cylinder of 10-m diameter and 10-m height, a slightly larger volume than used in the DoubleChooz experiment, filled with the Gd-loaded liquid scintillator.}
  \label{fig:geo_dc}
  \end{center}
\end{figure}

In a monolithic detector such as \dc, the prompt and delayed events take place in a single, continuous volume which is large compared to the distance traveled by the neutron. %
The positions of the positron and neutron events are then reconstructed using the relative time and / or charge registered by \pmts\ (omitted from the diagrams below for clarity), which are spread over the surface of the volume. 
The version of this geometry that we implemented in our \mc\ simulation is shown in Fig.~\ref{fig:geo_dc}. %
Note that it is considerably larger than the tanks used in the actual \dc\ experiment. %

\subsection{\label{subsec:seg}Segmented 3D: NuLat}

A key limitation of the monolithic approach is that its spatial resolution is typically larger than the neutron displacement needed to infer the source direction.
This means that the most immediate factor limiting a monolithic detector's directional capability is its spatial resolution.
The basic concept behind the \textbf{segmented} approach is to improve the spatial resolution by dividing the target volume into a lattice of cells which are each smaller than the desired distance measurement.

\begin{figure}[ht]
  \begin{center}
    \begin{overpic}[width=0.7\linewidth]{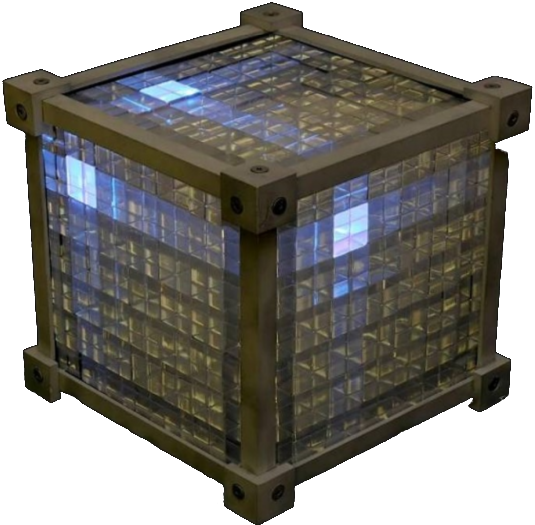}
    \end{overpic}
      \caption{Photograph of an ROL $7\times7\times7$ 
      demonstrator made of 1-inch cubes, with TIR behavior visible. An LED is placed inside one of the acrylic cubes, and its light is channeled along the three primary axes, allowing for easy identification of the originating cell~\cite{Lane:2015alq}. (Note: These cubes are about half the the size of the cubes used in NuLat itself and in our simulations, hence the similar total size of the $3^3$ array and the one shown here).}
      \label{fig:rol}
  \end{center}
\end{figure}

For our example segmented design, we chose the \nl\ detector~\cite{Lane:2015alq}, in which the target volume is a cube subdivided into cubical cells.
Interaction positions can then be determined to a resolution corresponding to the size of the cells.
The trick then lies in identifying which cell is active at a particular time.
\nl\ accomplishes this by employing \tir\ to create what is known as a \rol.
The \tir\ is made possible by a 1-mm air gap between scintillator segments.
Fig.~\ref{fig:rol} demonstrates the operating principle of an \rol: Light is channeled along each of the three principal axes from the originating cell to the outer faces, which would be instrumented with \pmts.
Combining the information from each face reveals the $\{x,y,z\}$ of the originating cell.

\begin{figure}
  \begin{center}
      \includegraphics[width=1.0\linewidth]{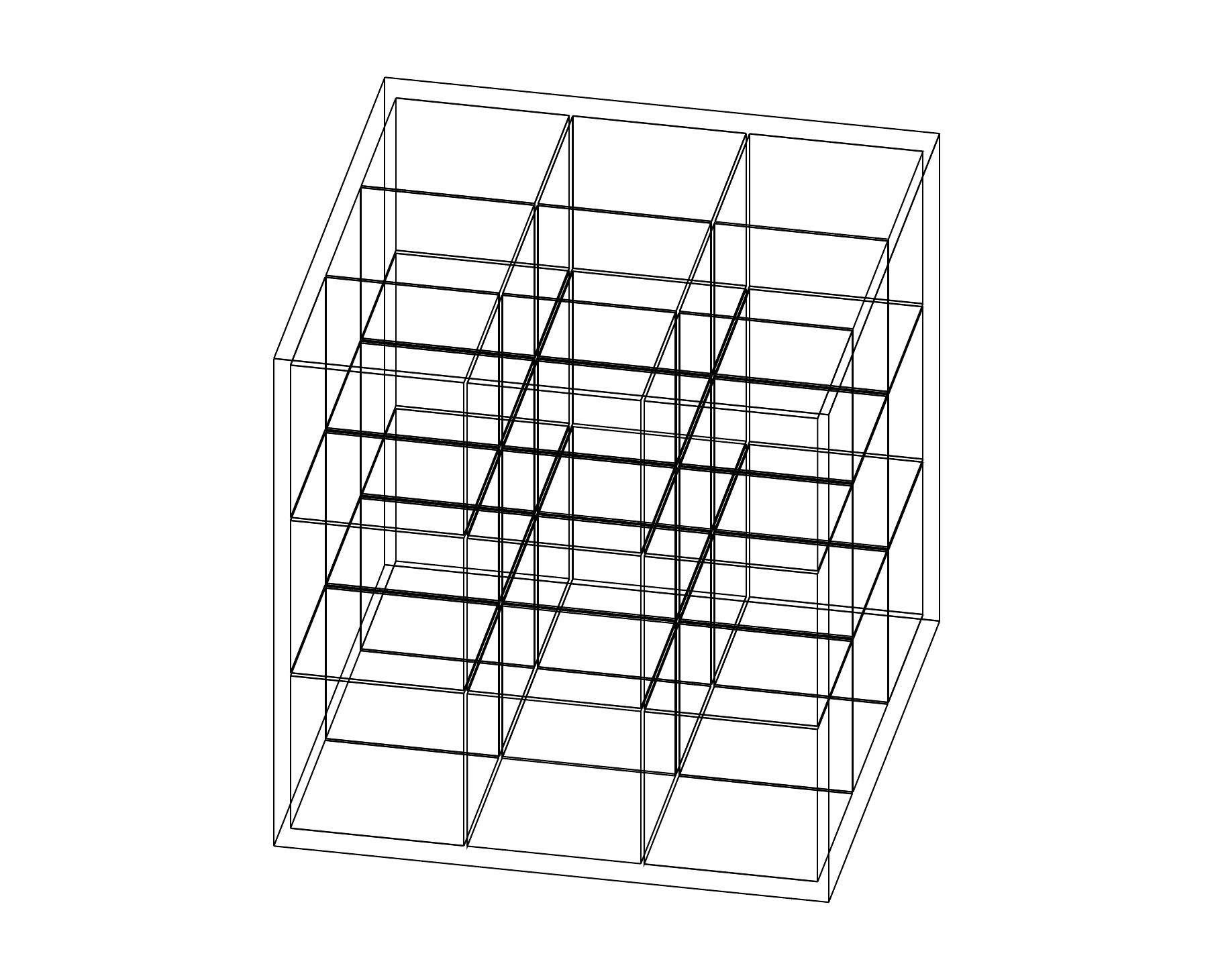}
      \caption{Simulated NuLat $3^3$ Geometry --- A $3\times3\times3$ array of $^6$Li-doped scintillator cubes. The cubes have a 5-cm side length and are separated by a 1-mm air gap. For this arrangement, the \ibd\ interaction vertices (and therefore the prompt events) are restricted to the central cube only; but the delayed events can occur in any of the 27 cubes.}
      \label{fig:nl_geo}
  \end{center}
\end{figure}

Directional information from \ibd\ interactions comes from events in which the prompt and delayed events occur in different cells; \rec\ is then simply a vector from the center of the first cell to the center of the second.
This also indicates that the cell dimensions should be comparable to the typical neutron travel distance ($d_n$) in the target material.
For this study, the cells were 5-cm cubes, comparable to the few-cm capture distances typical of our chosen materials.
The simplest configuration, $3 \times 3 \times 3$, gives 1 central cell with a neighbor in every direction; this is the configuration used in our \mc\ simulations and is shown in Fig.~\ref{fig:nl_geo}.
The \nl\ demonstrator~\cite{nulatThesis} that was built was $5 \times 5 \times 5$, and the actual NuLat is planned for $15 \times 15 \times 15$~\cite{Lane:2015alq}.\\

\subsection{\label{subsec:SANDD}Segmented 2D: SANDD}

\begin{figure}
\centering
      \includegraphics[width=1.0\linewidth]{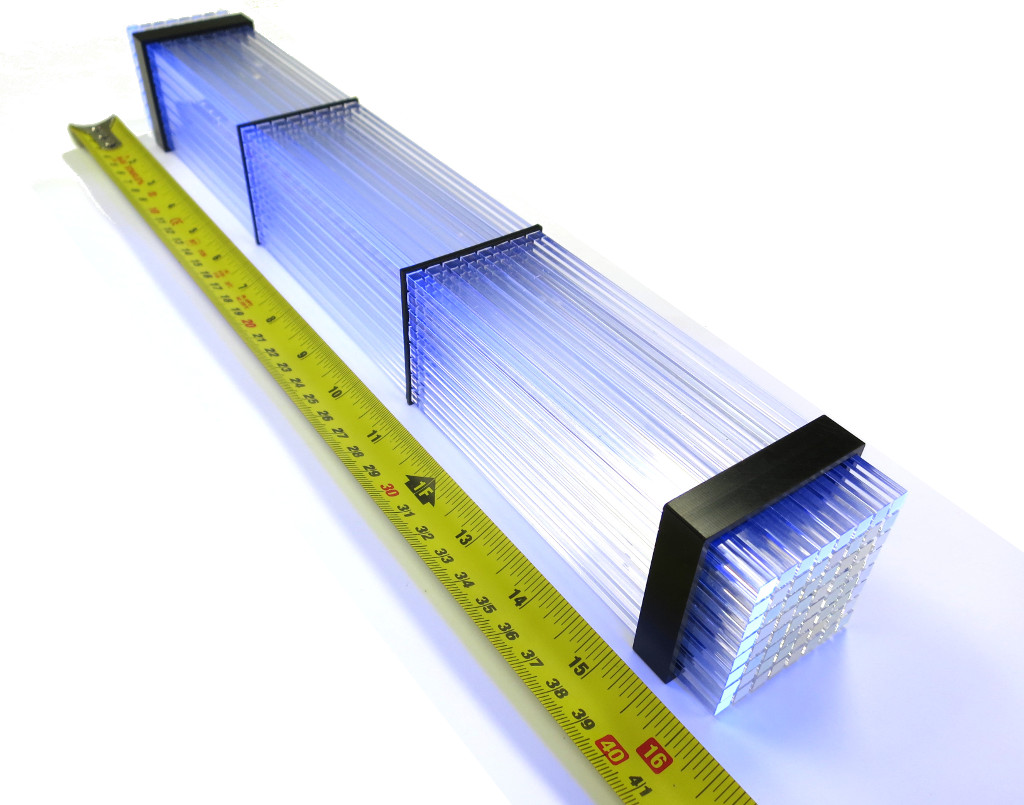}
      \caption{Photograph of the SANDD central module illuminated with an ultraviolet light --- an $8\times8$ bundle of square rods of $\nucl{6}{Li}$-doped pulse-shape-sensitive scintillator (5.4~mm $\times$ 5.4~mm $\times$ 40.64~cm, separated by a 1-mm  air gap with four alignment frames shown in black).}
      \label{fig:sandd_photo}
\end{figure}

Another recent segmented detector design is \sandd, a small (6.4-liter) array of $^6$Li-doped plastic scintillator with \psd\ capability~\cite{Sutanto_2021}.
(\psd\ makes it possible to distinguish certain particle types based on the shape of their scintillation pulses).
A prototype was designed and built for exploring near-field reactor monitoring, including sensitivity to IBD antineutrino direction.
The core concept is an $8\times8$ bundle of square scintillating rods ($5.4~\mathrm{mm} \times 5.4~\mathrm{mm} \times 40.64~\mathrm{cm}$) to detect both the initial interaction and the neutron capture.
This inner core of rods (hereafter referred to as the SANDD central module, or ``SANDD-CM'') is pictured in Fig.~\ref{fig:sandd_photo}.
It is designed to be surrounded by a single layer of square 2.54-cm scintillating bar segments and an additional layer of scintillating slabs ($2.54~\mathrm{cm} \times 5.08~\mathrm{cm}$, not pictured).
The configuration is quite similar to a Double Scatter Neutrino Camera built and tested at UH, as part of the program to study a Single Volume Scatter Camera for neutron direction determination, hosted by Livermore / SANDIA~\cite{Keefe_2022}.
We estimated that this prototype would have needed enlarging to ton-scale fiducial volume to have enough sensitivity for useful observations near a reactor, but it did show promising performance in lab tests.

The \nl\ and \sandd\ prototypes have demonstrated the viability of utilizing an array of scintillating parallelpipeds for \ibd\ detection, but they have also highlighted the problem of requiring many detector and electronics channels ($\sim10^4$ or more) when scaling up to the needed fiducial volumes for reactor studies, even from a few meters' distance. 

\subsection{\label{subsec:sep}Separation}

\subsubsection{\label{subsubsec:scat}The Neutron-Scattering Problem}

\begin{figure*}[ht]
  \centering
    \includegraphics[width=0.24\textwidth]{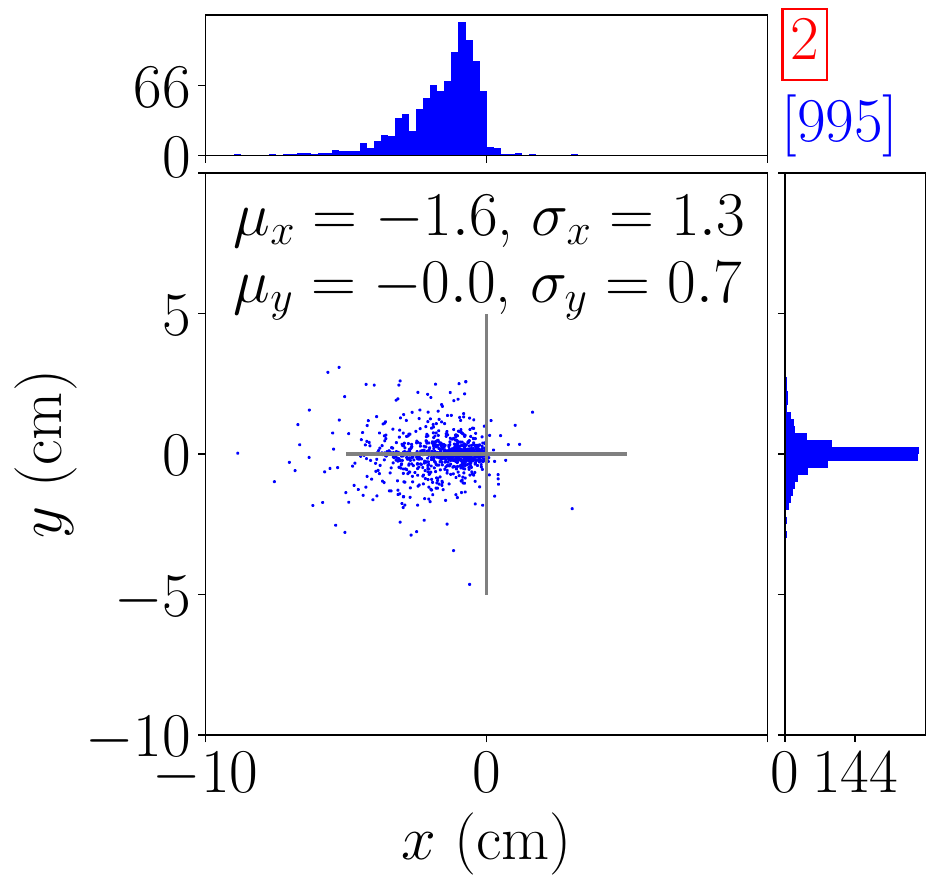}
    \includegraphics[width=0.24\textwidth]{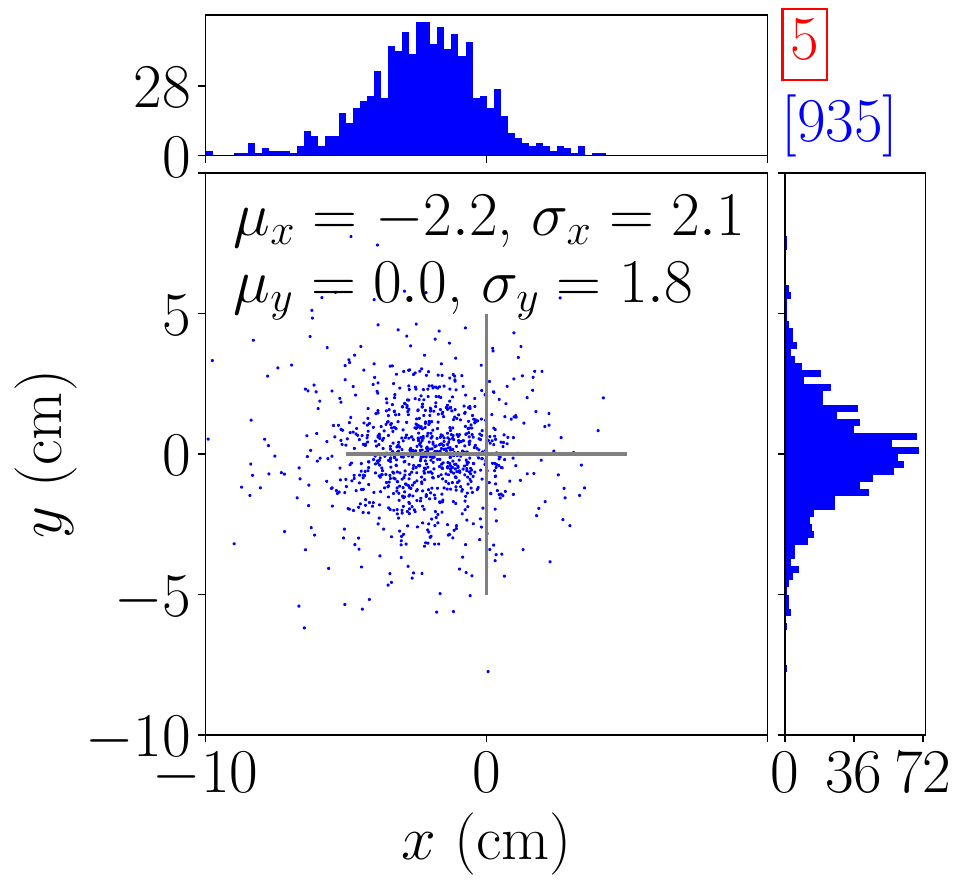}
    \includegraphics[width=0.24\textwidth]{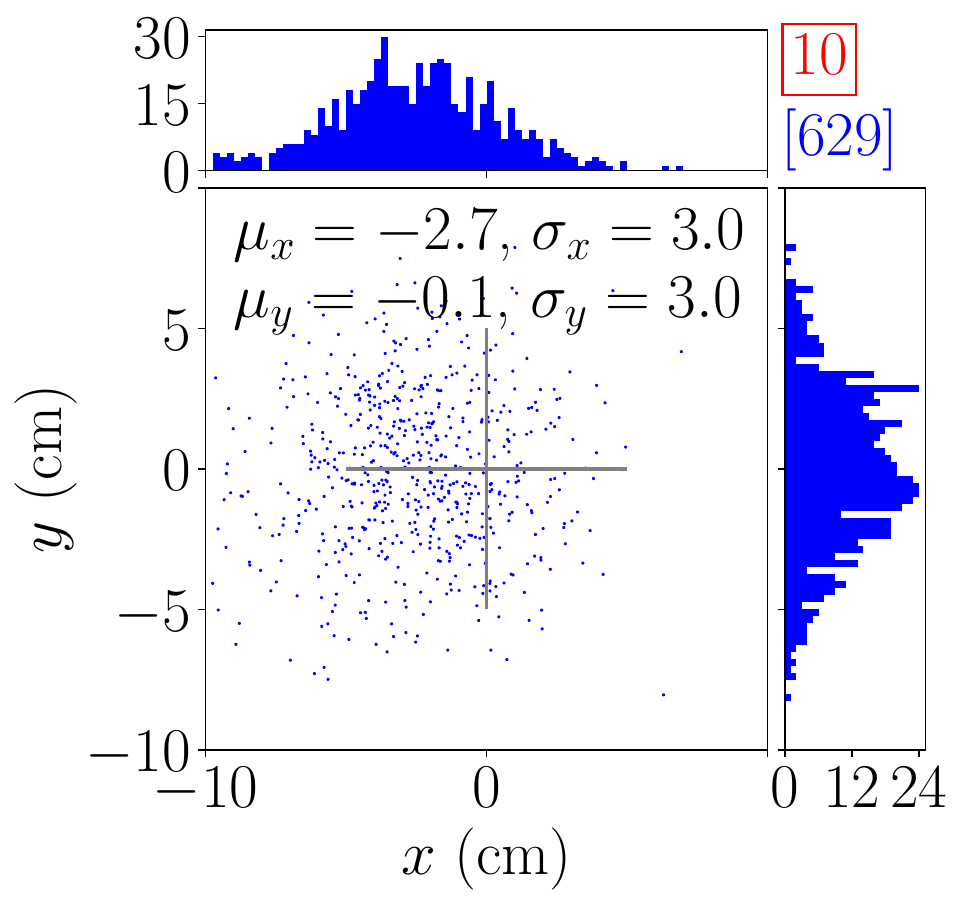}
    \includegraphics[width=0.24\textwidth]{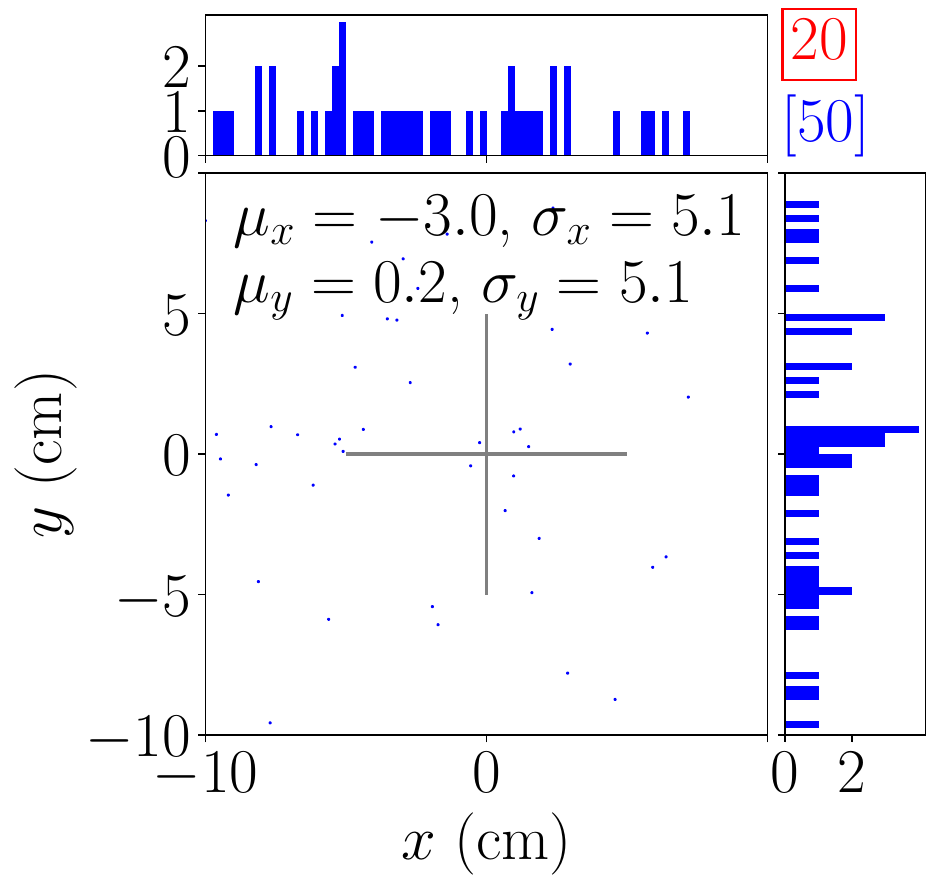}
    \includegraphics[width=0.24\textwidth]{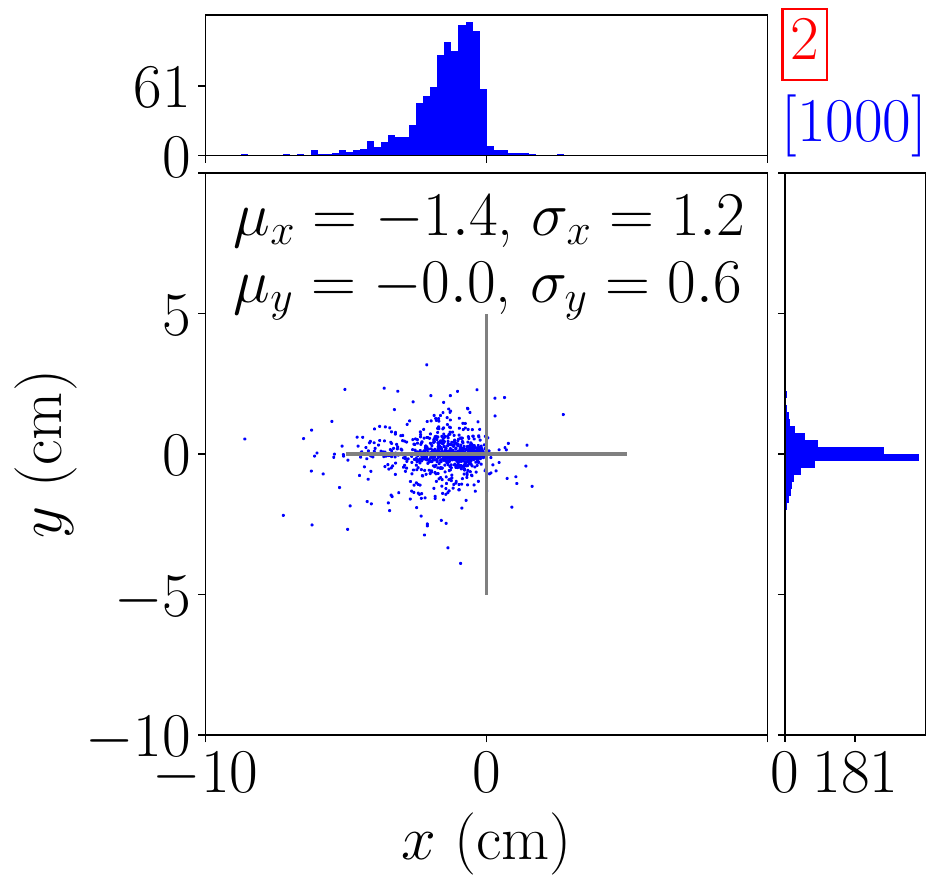}
    \includegraphics[width=0.24\textwidth]{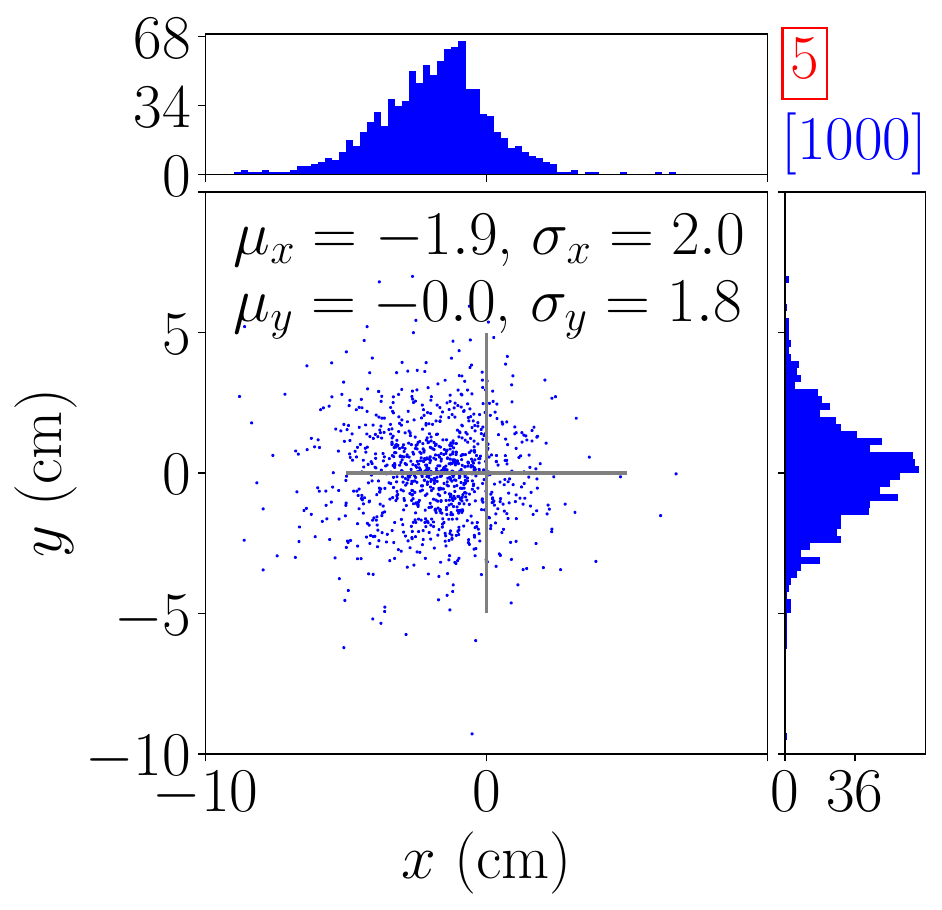}
    \includegraphics[width=0.24\textwidth]{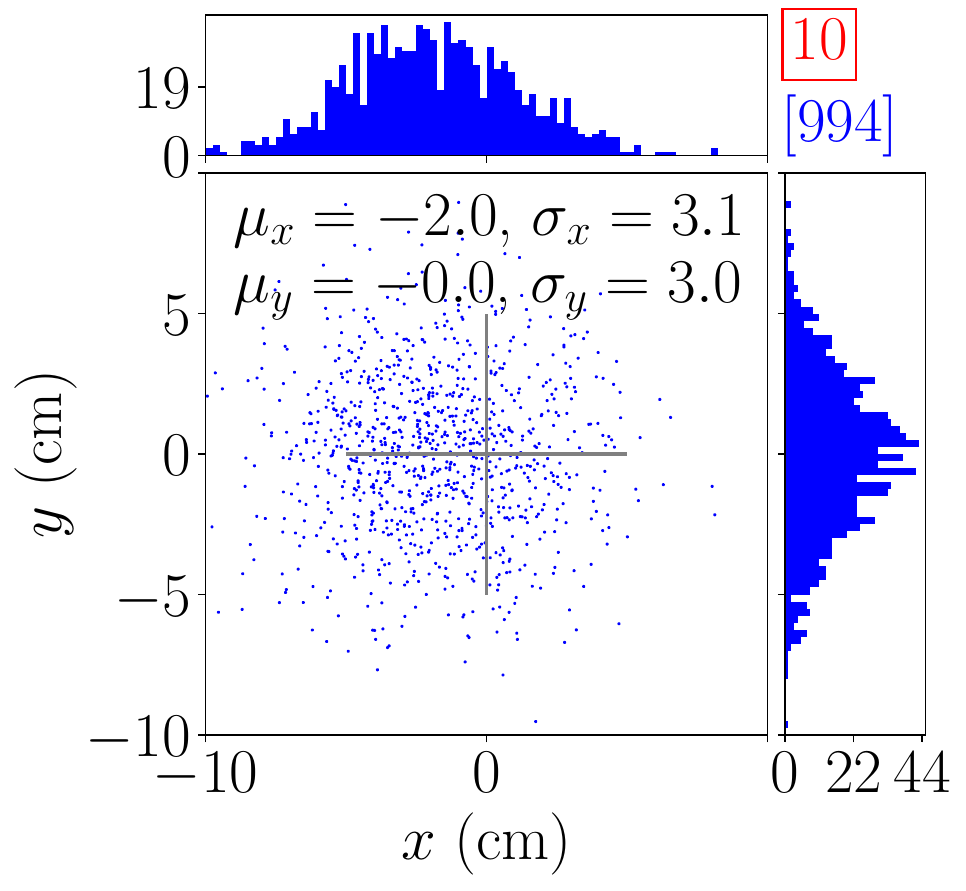}
    \includegraphics[width=0.24\textwidth]{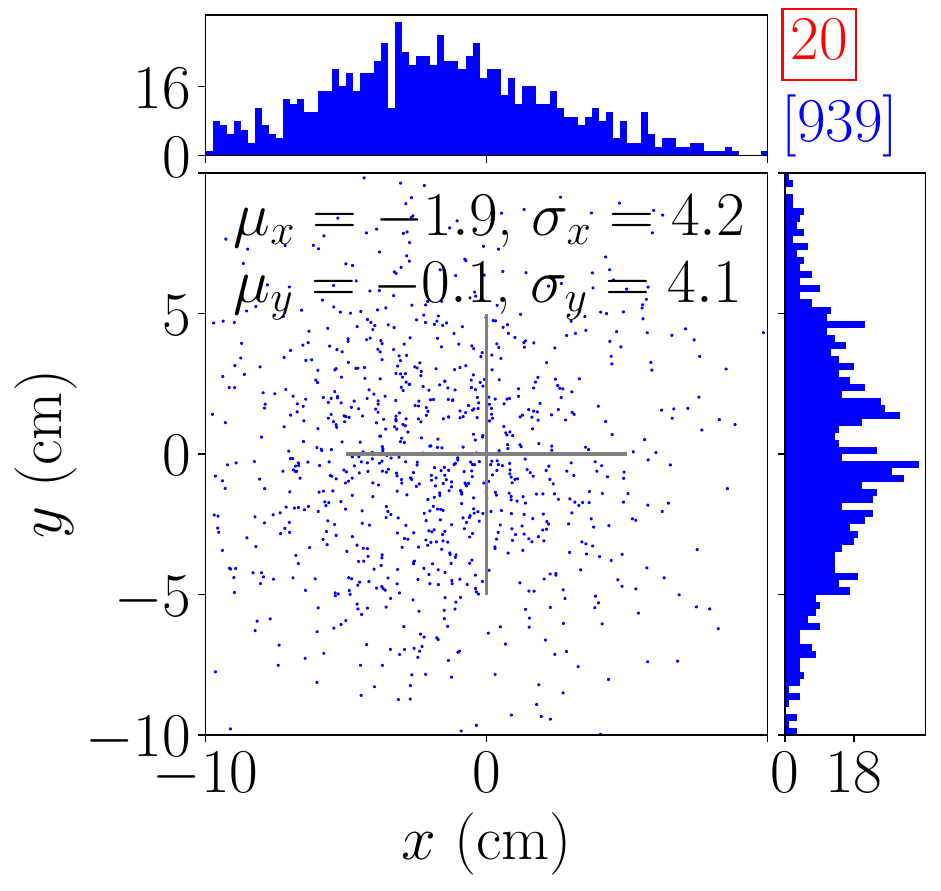}

  \caption{A set of 4-keV neutrons directed along \{-1,0,0\} were consecutively generated inside a large volume of plastic scintillator --- doped at 1.5\%wt $\nucl{6}{Li}$ ({\it top row of plots}) and undoped ({\it bottom row}); the directional effect of the scattering they undergo prior to capture is shown after 2, 5, 10, and 20 scatters (a typical total for this material). Displacements in the $xy$-plane; the gray cross (10-cm across) indicates the origin and the scale to make the displacement in the distributions more visible. For the undoped scintillator (bottom row), the neutrons undergo a higher number of scatters on average, making the displacement more obvious. }
  \label{fig:santa_scatbad}
\end{figure*}

\begin{figure}
\centering
\includegraphics[width=1.0\linewidth]{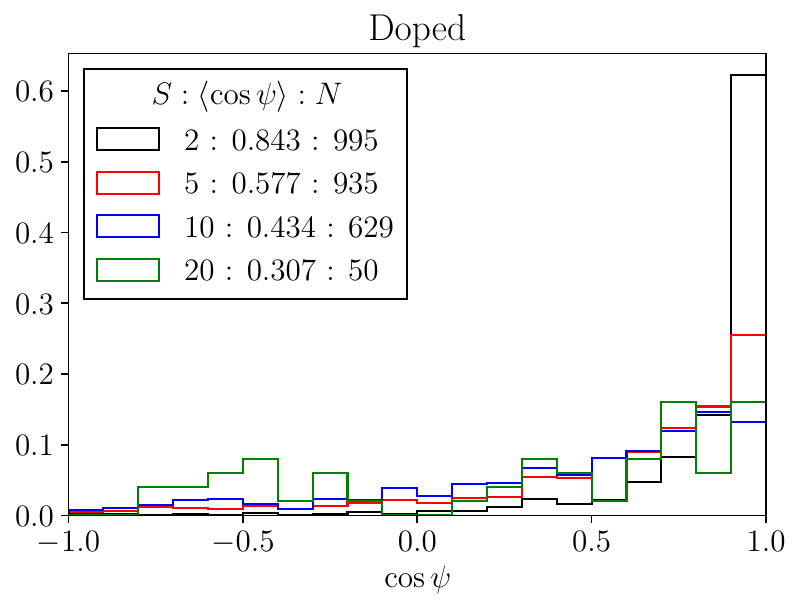}
\includegraphics[width=1.0\linewidth]{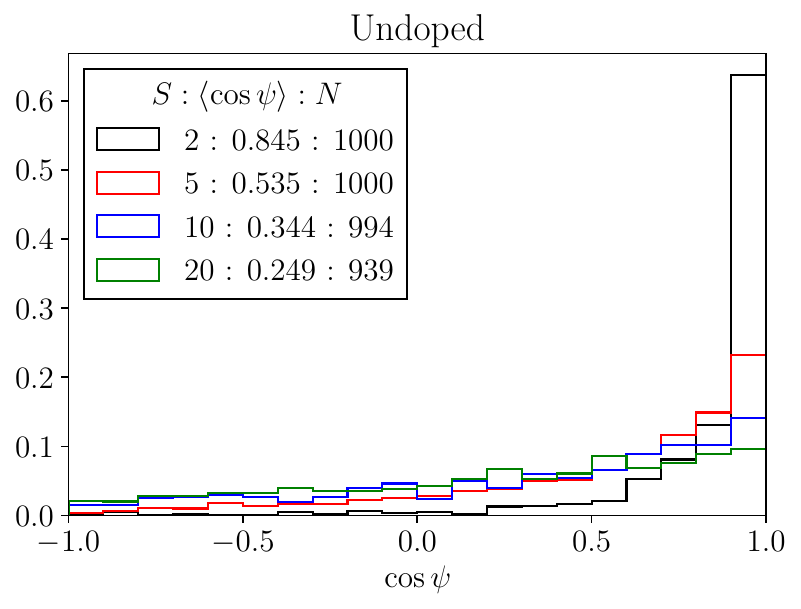}
\caption{Distribution of the $\cos \psi$ for various number of scatters (indicated in the legend along the mean values). The mean values indicate the strength of the directional trend. Number of neutrons left after $S$ number of scatters is indicated in the legend, i.e. the number of entries in the corresponding histogram.}
\label{fig:cosPsi_monoE_neutrons}
\end{figure}

A fundamental challenge for any \ibd\ detector is that neutron scattering quickly degrades directionality.
IBD neutrons do not have a uniform initial direction (even if the incident antineutrinos do), so they were not well-suited to illustrate the directionality effect at the IBD energies. %
In order to show {\it only} the directional effects due to neutron scattering, %
we simulated a uniform set of 1000 neutrons with kinetic energy of $4~\mathrm{keV}$ in a piece of plastic scintillator doped at 1.5\%wt $\nucl{6}{Li}$.
Fig.~\ref{fig:santa_scatbad} shows relevant parameters for these neutrons after 2, 5, 10, and 20 scatters.
It is worth noting that all of these parameters \dash\ most notably the width of the spread and the total number of scatters \dash\ can vary substantially depending on the scintillator material, choice of dopant, and doping concentration.
The raw Monte Carlo positions are shown in Fig.~\ref{fig:santa_scatbad}a.
After the first scatter, the distribution is quite clearly forward-biased, but it spreads out considerably in just a few more scatters.
After 20 scatters \dash a typical total in this material \dash the distribution is only slightly heavier in the forward direction than in the backward direction, and the transverse spread is comparable to the forward-backward spread.
Fig.~\ref{fig:santa_scatbad}b shows a quantitative view of this process, representing the overall directional information via the distribution  of $\cos \psi $ at each step.
In $N=2$, the momenta are strongly concentrated in a narrow cone about the initial momentum, but in $N=5$, the cone is considerably broader and much less sharply-defined.
The stark contrast between $N=2$ and $N=5$ indicates how detrimental even a few scatters are to the directional information carried by the neutron.
By the time the neutron captures at around $N=20$, the trend in $\cos \psi $ is relatively weak.
In a real detector with finite position resolution \dash\ as opposed to the \mc-truth positions shown here \dash\ the reconstructed trend is weaker still.

The problem this causes in reconstructing the direction of the original incident \nb\ is that the position distributions in $\{x,y,z\}$ of the prompt and delayed events become a pair of closely-overlapping Gaussians, both having standard deviations that are considerably larger than the separation between their means.
This is another reason (in addition to the stochasticity discussed in \S\ref{subsec:dirn_stats}) that many events are needed to resolve the source direction.
This also indicates that improving the position resolution $P$ of a traditional \ibd\ detector can only increase its directional capability up to the point at which $P \sim d_{n}$, which is $\mathcal{O}(1~\mathrm{cm})$ in the relevant materials.
At this point, the neutron-scattering process becomes the dominating factor limiting improvements in the detector's angular resolution.

\subsubsection{\label{subsubsec:santa_sln}The SANTA Solution}

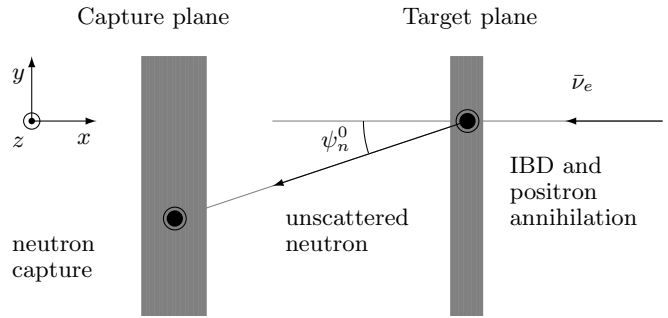
\begin{figure}[ht]
\setlength{\unitlength}{.1\linewidth}
\begin{picture}(10,5)
  \thicklines
        \multiput(2,0)(.025,0){40}{\color{gray}\line(0,1){4}} %
        \multiput(6.75,0)(.025,0){20}{\color{gray}\line(0,1){4}} %
  \thinlines
\put(7,3){\color{gray}\line(-4.5,-1.5){4.5}} %
\put(7,3){\vector(-4.5,-1.5){3}} %
 \put(1., 4.5){\text{Capture plane}}
  \put(6, 4.5){\text{Target plane}}

  \put(10,3){\color{gray}\line(-1,0){6}} %
  \put(10,3){\vector(-1,0){1.5}} %

  \put(.3,3){\vector(1,0){1}} %
  \put(.3,3){\vector(0,1){1}} %
  \put(.3,3){\circle{0.25}} %
  \put(.3,3){\color{black}\circle*{0.1}}
  \put(1., 2.65){$x$}
  \put(0., 3.65){$y$}
  \put(0., 2.6){$z$}
  \put(7,3){\color{black}\circle*{.25}} %
  \put(2.5,1.5){\color{black}\circle*{.25}} %
  \put(7,3){\color{black}\circle{.35}} %
  \put(2.5,1.5){\color{black}\circle{.35}} %

  \put(8.6, 3.5){$\bar\nu_e$}
  \put(7.65, 2.2){\text{IBD and}}
  \put(7.65, 1.8){\text{positron}}
  \put(7.65,1.4){\text{annihilation}}
  \put(0,.6){\text{capture}}
  \put(0,1){\text{neutron}}

  \put(4.2,1.4){\text{unscattered}}
  \put(4.2,1.0){\text{neutron}}

  \put(7,3){\arc[180,198]{1.6}} %
  \put(4.75, 2.6){$\psi_n^0$}

\end{picture}
        \caption{Design concept of a target-capture-plane detector based on the SANTA. In this study, the antineutrino direction was perpendicular to the detector planes. Only one capture plane was implemented (unlike the original design with the second capture plane). Angle $\psi_n^0$ is the angle between the direction of capture-IBD cells and the $x$-axis. In the separation detector like SANTA, the mean values for angles $\psi_n^0$ and $\varphi$ are the same.}

\label{fig:santa_concept}
\end{figure}

The loss of directional information due to neutron scattering is an inherent limitation in bulk-volume detectors, whether they are monolithic like \dc\ or subdivided like \nl. %
One proposed solution to this problem is the \santa~\cite{Safdi:2014hwa}, which allows the neutron to travel much further in its original direction before its first scatter.
This requires the neutron to leave the target material almost immediately after it is produced, then travel some relatively long distance, then enter a second volume of target material and capture there.
As shown in Fig.~\ref{fig:santa_concept}, \santa\ proposes to achieve this by arranging the scintillator materials into two separate layers: a very thin ``target'' layer for the prompt event and neutron production, and a thicker ``capture'' layer for neutron thermalization and capture.
In order to encourage the neutron to capture in the correct layer, the target layer would not be doped with any neutron-capture agents, and the capture layer would be doped rather heavily --- for example, polyvinyl toluene (PVT) at 5\%wt natural boron in the original proposal.

The dramatic benefit of this approach is that neutron scattering has only a minimal effect on the reconstructed source direction.
This can be seen in Fig.~\ref{fig:santa_concept}:
In most \ibd\ detectors, the neutron-capture region surrounds the \ibd\ vertex;%
and the randomness of the neutron's scattering almost completely determines the angle $\varphi$.
\santa\ sidesteps this problem by moving the scattering downstream, so that \nnc\ captures anywhere within the scattering region will point approximately along the true source direction.
We refer to this $target$-$segment \rightarrow flight \rightarrow capture$-$segment$ arrangement as a \textbf{separation} detector.

It is worth noting that the neutron is not guaranteed to capture in the capture layer, even with heavy doping.
A neutron may, for example, bounce off of the capture layer and eventually capture back in the target layer, possibly reflecting off of nearby walls or shielding in the process.
(Such captures typically occur on an $\nucl{1}{H}$ in the plastic).
These relatively-rare events could still be used for measuring the total antineutrino rate but would generally be cut from the directional analysis.

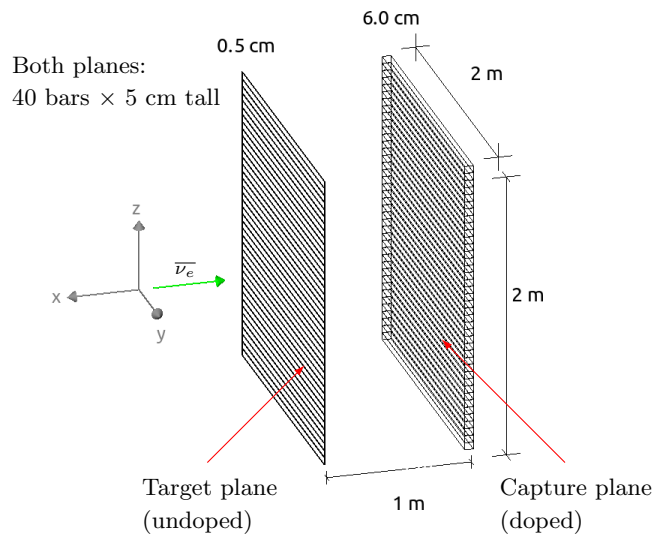
\begin{figure}[ht]
  \begin{center}
    \begin{overpic}[width=1.0\linewidth]%
      {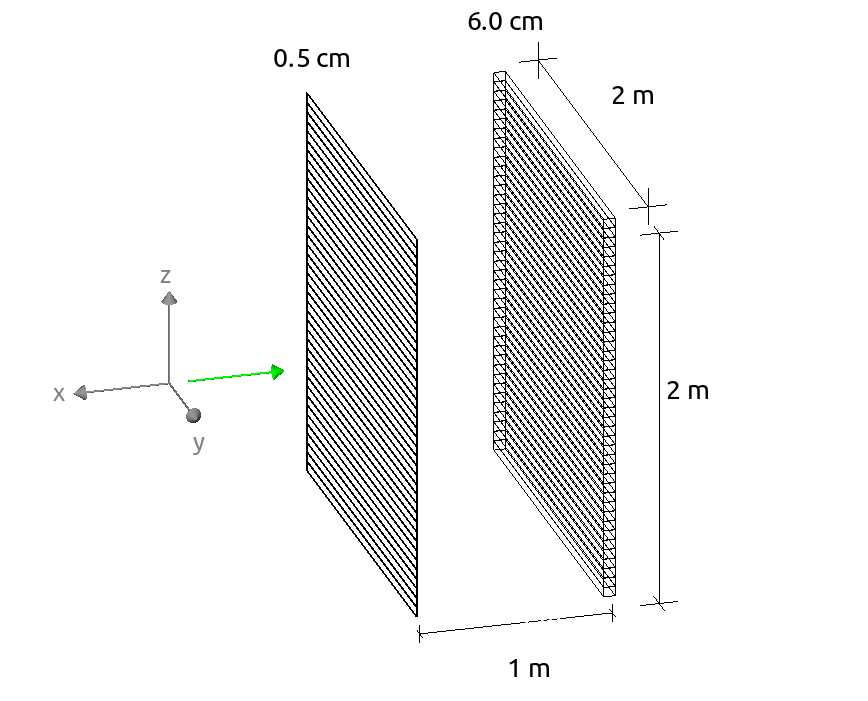}
      \put(20,5){Target plane}
      \put(20,0){(undoped)}
        \put(30,10){\color{red}\vector(1,1){15}}
      \put(0,70){Both planes:}
      \put(0,65){40 bars $\times$ 5~cm tall}
      \put(75,5){Capture plane}
      \put(75,0){(doped)}
      \put(85,10){\color{red}\vector(-1,1){19}}

      \put(25,39){\scriptsize \nbc}
    \end{overpic}
  \caption{Simulated \santa\ Geometry --- the ``target'' plane comprised of 40 bars of undoped scintillator 2~m~$\times$~0.05~m~$\times$~0.005~m; the ``capture'' plane comprised of 40 scintillator bars doped with natural boron, 2~m~$\times$~0.05~m~$\times$~0.06~m. The separation between the two planes is 1~m, resulting in a ``packaging factor'' of about 0.46\% (the ratio of the target volume to the total volume of the detector).}
  \label{fig:santa_geo}
  \end{center}
\end{figure}

The implementation of \santa\ used in this MC study was based on a possible realization proposed in~\cite{Safdi:2014hwa}.
It contains a 0.5-cm-thick undoped target layer separated by 1~m from a 6-cm-thick capture layer doped at 5\%wt~natural~boron, as shown in Fig.~\ref{fig:santa_geo}.
Both layers are composed of PVT plastic scintillator and are 2~m square.
The original proposal included an additional layer, upstream of the target layer, for catching positrons that escape the target layer.
We elected to forego this feature, both to keep the configuration minimal and because the energy deposition from positron tracks in the target layer was sufficient for our purposes.
No explicit mechanism was proposed by~\cite{Safdi:2014hwa} for measuring the locations of the interactions in the scintillator.
At \uh, we took this concept and considered how it might be realized in an actual detector.
To this end, we subdivided each plane into a set of 40 horizontal scintillator bars, each 50~mm tall and separated by 1~mm from the next bar.
We inserted a 0.5-mm strip of black acrylic for optical separation between each bar.
In a real detector, a 5-cm \pmt\ or \sipm\ would then be placed at either end of each bar, for 160 light sensors total.
For a typical prompt or delayed event, a strong signal will be produced in exactly one bar.
The event's position can then be extracted as follows:
The vertical component $(z)$ is taken to be the value of $z$ at the center of the bar.
The axial component $(x)$ is taken to be the value of $x$ at the center of the plane (so, in our implementation, $x~=~\pm50~\mathrm{cm}$).
The transverse component $(y)$ is determined by comparing light output at either end of the bar.
If fast-timing measurements are available, the transverse determination could be enhanced by comparing photon arrival times at either end of the bar.

Finally, \santa\ is unique among the designs studied here in that it does not have a full \fov\ \dash\ that is, it is sensitive only to antineutrinos coming generally from its ``forward'' direction ($+x$ in Figs.~\ref{fig:santa_scatbad}a,~\ref{fig:santa_concept}, and~\ref{fig:santa_geo}.
This does bias the results: \santa\ will always report a source direction within its \fov.
We have run preliminary simulations using off-axis neutron beams; the results suggest that, after accounting for a systematic skew towards the center, \santa\ can reliably identify off-axis sources within its \fov.
More specifically, \santa's \fov\ is approximately a cone (3D) or arc (2D) of half-angle $45^\circ$.
This is $\sim15\%$ of $4\pi~\mathrm{sr}$ in 3D and exactly 25\% of $2\pi~\mathrm{rad}$ in 2D.
Whether this limited \fov\ is a drawback or a benefit depends on the objective of a given deployment.
For example, it is clearly detrimental if attempting to locate an unknown reactor with an arbitrary position; but it can suppress noise from irrelevant directions if attempting detailed measurements near a known reactor.

\subsection{\label{subsec:hyb}Hybrid}

\subsubsection{\label{subsubsec:hyb_mot}Motivation}

Our collaboration felt that the \santa\ approach featured an ingenious insight on solving the aforementioned neutron-scattering problem; however, we also anticipated some major practical limitations with the design as laid out in the original paper. %
Our two biggest concerns were that, relative to the more traditional designs:
\begin{enumerate}
  \item The much-larger surface-area-to-volume ratio would expose the detector to severe levels of backgrounds, especially in our targeted near-surface deployment scenario.
  \item The much-smaller target-volume-to-footprint ratio would severely restrict the achievable target mass in a real deployment environment.
\end{enumerate}
It is worth noting that the \santa\ design does inherently compensate somewhat for both of these effects, in that it requires many fewer events to achieve a given angular resolution.
However, our simulations indicated that this compensation alone would likely be insufficient to make the original design practically viable. %

We therefore set out to devise a class of detector designs that would strike a balance between \santa's directional performance and the more-traditional detectors' larger target mass and stronger resilience against backgrounds.
In this work, we refer to such detector plans as \textbf{hybrid} designs.
All of them seek to improve directional performance by incorporating space for the neutron to exit the target medium almost immediately after production and then undergo thermalization and capture in a different target segment, which is the driving behavior behind \santa's directional capabilities.

\subsubsection{\label{subsubsec:hyb_chkbd}``Checkerboarding''}

One of our primary approaches was to provide this free-flight space by simply removing alternating detector segments from the \sandd\  and \nl~$15^3$ configurations in a ``checkerboard'' pattern.
This pattern is determined by the row, column, and (in 3D) layer numbers for any given cell in the lattice.
There are two types of cells in this arrangement:
\begin{enumerate}
  \item \textit{Active} cells, which are occupied by a scintillator segment as usual.
  \item \textit{Inert} cells, which are either left empty (air, vacuum, etc.) or, if necessary, are filled with some material which is effectively transparent to neutrons and (for some geometries) optical photons.
\end{enumerate}

\begin{figure}[ht]
    \includegraphics[width=1.\linewidth]{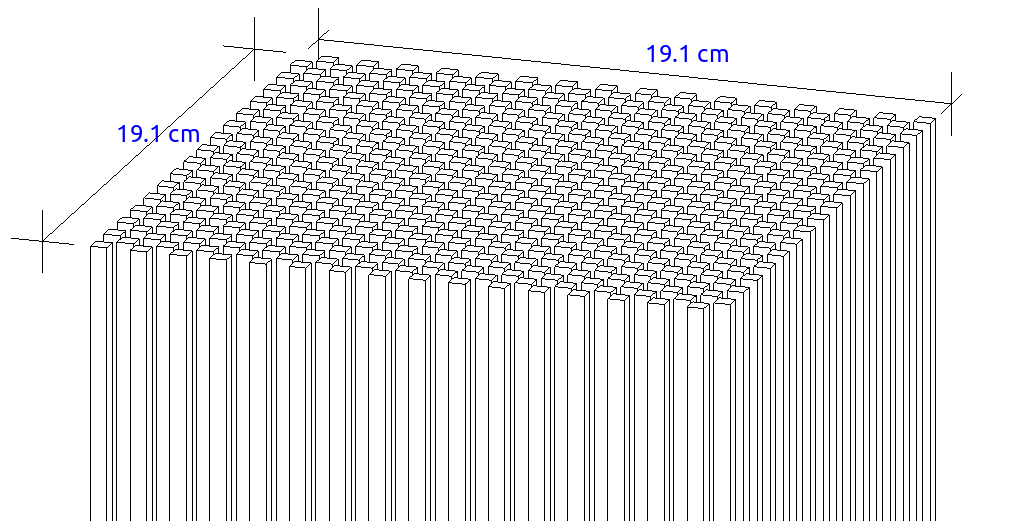}
    \textit{(a) Closeup showing lattice arrangement} 
    \includegraphics[width=1.\linewidth]{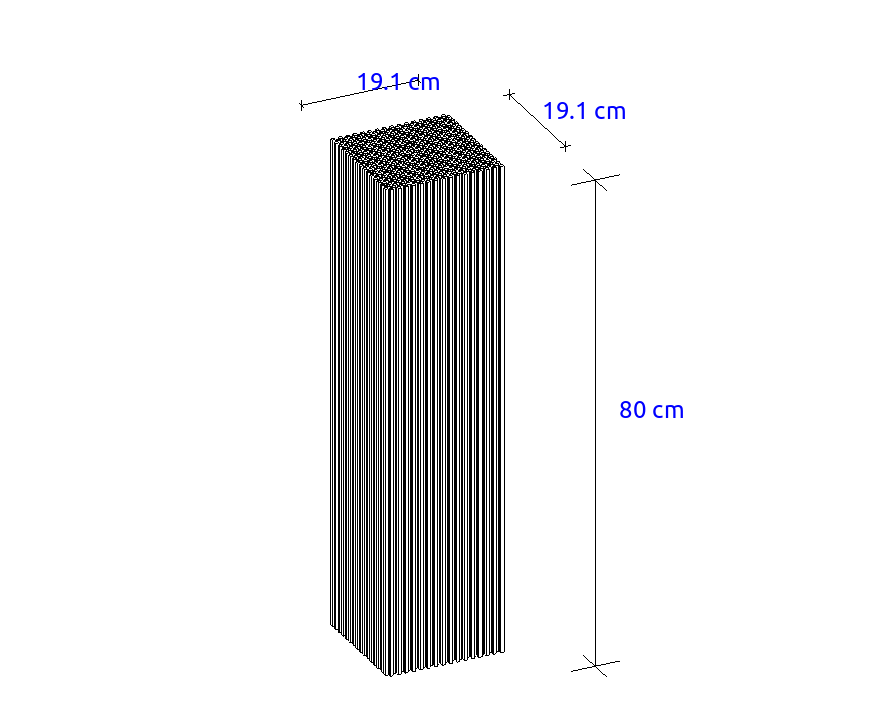}
    \textit{(b) Full array used in simulations.} 
    \caption{2D checkerboard geometry.  The lattice is 32 rods on a side. }
    \label{fig:chkbd_2d}
\end{figure}

\begin{figure}[ht]
    \includegraphics[width=1.\linewidth]{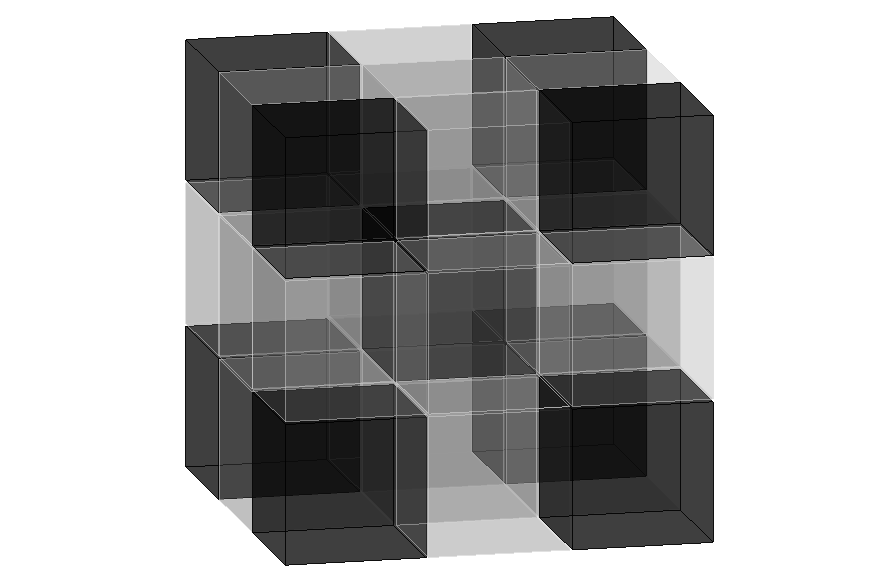}
    \textit{(a) Closeup showing lattice arrangement (inert cells ghosted)} 
    \includegraphics[width=1.\linewidth]{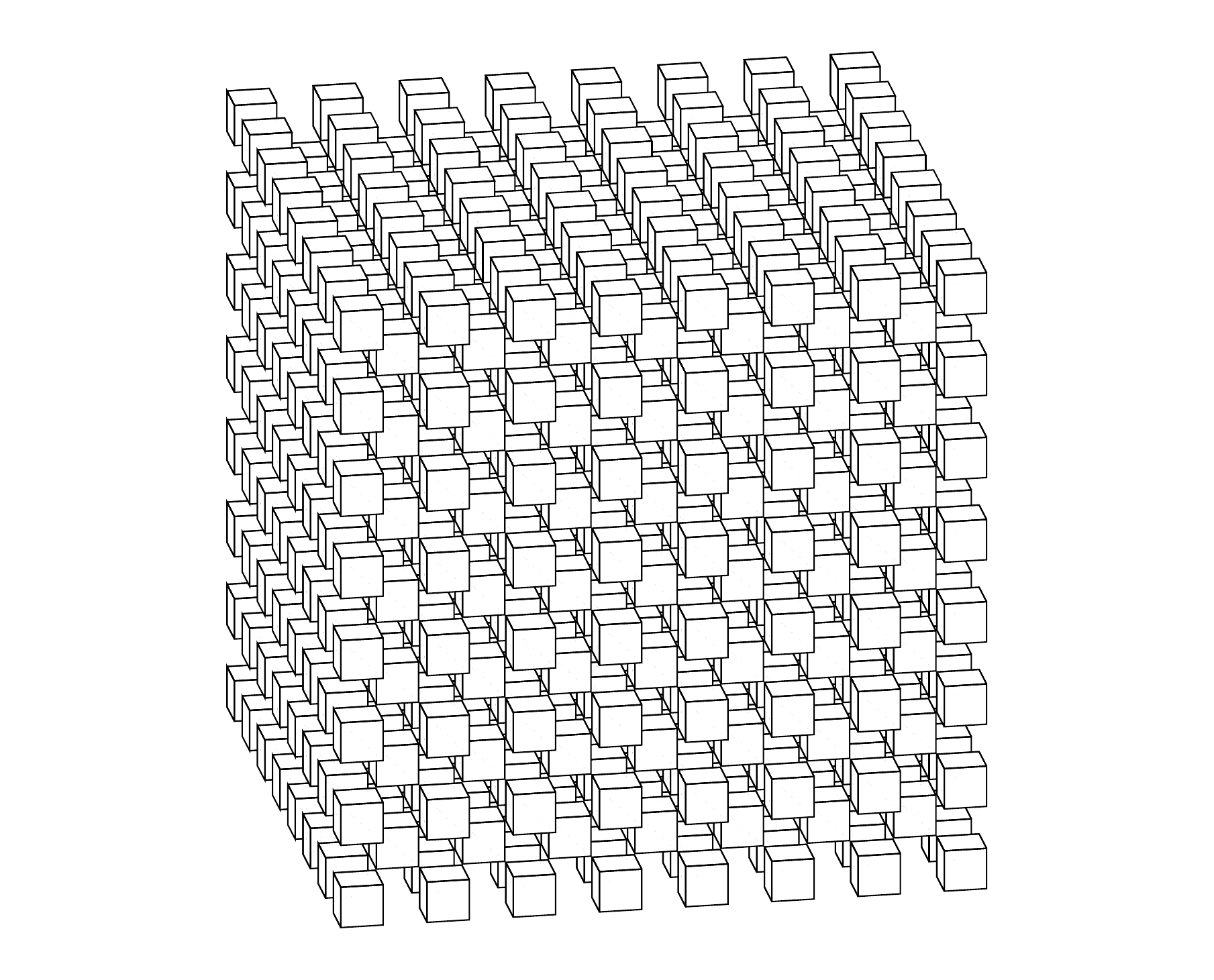}
    \textit{(b) Full array used in simulations.} 
    \caption{3D checkerboard geometry. The lattice is 15 cubes on a side. Each cube is 5-cm on a side. }
    \label{fig:chkbd_3d}
\end{figure}

Lattice cells satisfying Eq.~\ref{eq:chkbd} are made active, and all others are made inert:
\begin{equation}
  \label{eq:chkbd}
  \begin{split}
    \mathit{(2D)}\quad\ ROW\ \%\ 2\ &=\ COL\ \%\ 2 \\
    \mathit{(3D)}\quad\ ROW\ \%\ 2\ &=\ COL\ \%\ 2\ =\ LYR\ \%\ 2\ ,
  \end{split}
\end{equation}
where ``\%'' is the modulus operator.
In other words, a lattice cell is made active if and only if its row, column, and (in 3D) layer numbers are all either simultaneously even or simultaneously odd.
Renderings of our simulated versions of these arrangements are shown in Figs.~\ref{fig:chkbd_2d} and~\ref{fig:chkbd_3d}.
Note that the 3D version does not identically reproduce the traditional 2D-checkerboard pattern in any single plane.
Instead, it produces a pattern in which each active cube only has active neighbors at its vertices, rather than against its faces.
This is the three-dimensional analogue to the traditional 2D checkerboard, wherein black squares only have black neighbors at their corners rather than along their sides.
It is this ``neighbors on vertices only'' property that provides flight space for \ibd\ neutrons within an otherwise bulk-volume detector.

For this study's simulations, all inert cells were simply left empty (\textit{i.e.}, they contained only air).
We leave the investigation of candidate inert materials to future work, other than to note the following:
To qualify, a material's interaction lengths for both neutron elastic scattering and neutron capture should be at least comparable to (and ideally much larger than) the cell size: $\{\:\lambda_{es}\,,\:\lambda_{cap}\:\}~\gtrsim~\{l,w,h\}_{cell}$.
If the inert cells also need to serve as light guides, such as in the \nl\ / \rol\ arrangement, then the material must be sufficiently transparent to optical photons as well.
We have run preliminary simulations indicating that glass/quartz $(\mathrm{SiO}_2)$ may be one such material.

\section{\label{sec:method}Methodology}

We ran multiple sets of \mc\ simulations to compare the directional performance of these various detector geometries.
Because we wanted to specifically focus on the inherent potential directionality of each approach, we determined to \emph{not} include backgrounds at this stage.
The idea here \dash particularly for the \santa\ design \dash was to investigate whether the (expected) improvement in directional capability offered sufficient reason to proceed to addressing the (expected) difficulties with backgrounds.
Further details on each set are discussed below.

\subsection{\label{subsec:ncomp_params}Shared Parameters}

We used RAT-PAC~\cite{Seibert:2006}~\cite{Duvall:2022} (an implementation of GEANT4) and ROOT~\cite{Brun:1997}~\cite{Brun:2015} for simulation and analysis; it was adapted by various collaborations including SNO+~\cite{PhysRevLett.130.091801} and AIT~\cite{Li_Gd_water, PhysRevApplied.20.034073, PhysRevApplied.19.034060}.
We simulated 10~000 IBD events in each detector.
We took the \nb\ source to be distant enough that the incoming antineutrino flux could be well-represented by a plane wave; \textit{i.e.}, we took the incoming antineutrinos to be uniform in direction.
Furthermore, we took the \nb\ source to be located ``at the horizon'' relative to the detectors.
For convenience, we placed our origins at the center of each detector with the $x$-axis pointing toward the \nb\ source, so that the incoming antineutrinos were traveling along $\{x,y,z\} = \{-1,0,0\}$.
We then oriented each detector to be ``facing'' the \nb\ source \dash\ that is, we aligned each detector for its optimal directional sensitivity for the chosen antineutrino direction.
The energy distribution for the \ibd\ events was taken from the reactor spectrum built into \rat~\cite{Seibert:2006}.
The \ibd\ vertices were generated in a uniform (flat) random distribution throughout a volume or set of volumes chosen specifically for each detector.
IBD events were separated in time by a Poisson distribution of mean 1~s.
In all cases but NuLat $3^3$, IBD vertices were distributed uniformly throughout the scintillator volume(s).
In NuLat $3^3$, IBD vertices were constrained to the central cube.

Backgrounds were expressly excluded from this study and will be addressed in subsequent work. 
This allows us a clear examination of the potential directional capability inherent to each geometry, with the goal of assessing which designs we wish to investigate further (including studies of performance in the presence of backgrounds).
It also prevents our discussion from becoming needlessly restricted to a particular deployment setting, as background profiles are strongly site-dependent.

Realistic position resolution was simulated via the following:
\begin{enumerate}
    \item For axes with segmentation: Discretizing the relevant coordinate(s).
    \item For axes without segmentation: Adding noise to the raw MC data until matching the resolution of the respective experiment (\textit{e.g.,} 15~cm for the monolithic / Chooz-type experiment).
\end{enumerate}
Remaining parameters are listed with respective results.

During this process, we developed software tools for analysis and visualization that we hope will be useful to the community in general and to ROOT/\rat\ users in particular (see~\cite{Duvall:2022}).
The results of these simulations are presented in \S\ref{sec:results} below.

\subsubsection{\label{subsubsec:selection}Detector Selection}

Now that we have described each of the experiments we chose to simulate, we would like to briefly discuss why we selected this particular collection.

There are a number of experiments we could have chosen to represent each fundamental detector type.
The field in general is well-represented in the 2015 survey of planned short-baseline reactor-antineutrino experiments shown in Table~1 in NuLat whitepaper~\cite{Lane:2015alq}. %
Most of the detectors in the table were designed for a simple measurement of the antineutrino flux and not for antineutrino directionality.
As a result, our examination of one or two designs from each type \dash monolithic / minimally-segmented (\dc), 2D-segmented (PROSPECT, \sandd), and 3D-segmented (\nl) \dash covers the field of short-baseline reactor experiments as far as antineutrino directionality is concerned.

Because members of our group personally worked on the \mtc, \nl, and \sandd\ experiments, and our faculty includes members of the \dc\ and PROSPECT collaborations, these detectors were naturally optimal choices for use in this study.
The \santa\ design is the only one of its type, so its inclusion was essentially predetermined.
Finally, the hybrid designs appear here because they were developed by our group as a part of this work.

\section{\label{sec:results}Primary Results}

\subsection{\label{subsec:results_ex}Selected Examples}

\begin{figure*}[ht]
    \centering
    \begin{overpic}[width=0.32\linewidth]%
    {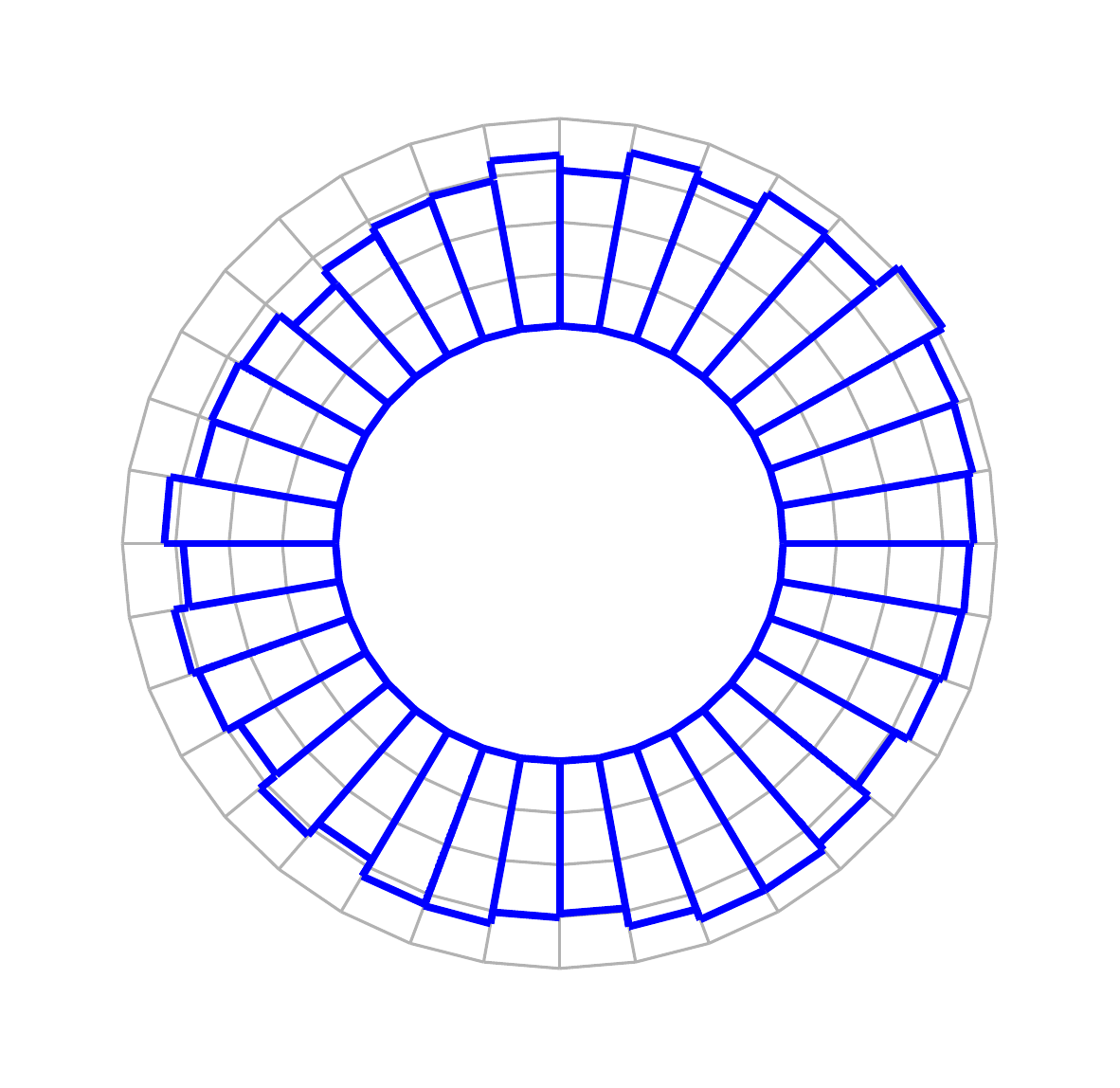}
        \put(91,47){\color{black}0}
        \put(46,90){\color{black}$\pi/2$}
        \put(5,47){\color{black}$\pi$}
        \put(45,5){\color{black}$3\pi/2$}
        \put(38,47){\color{black}Monolithic}

    \end{overpic}
    \begin{overpic}[width=0.32\linewidth]{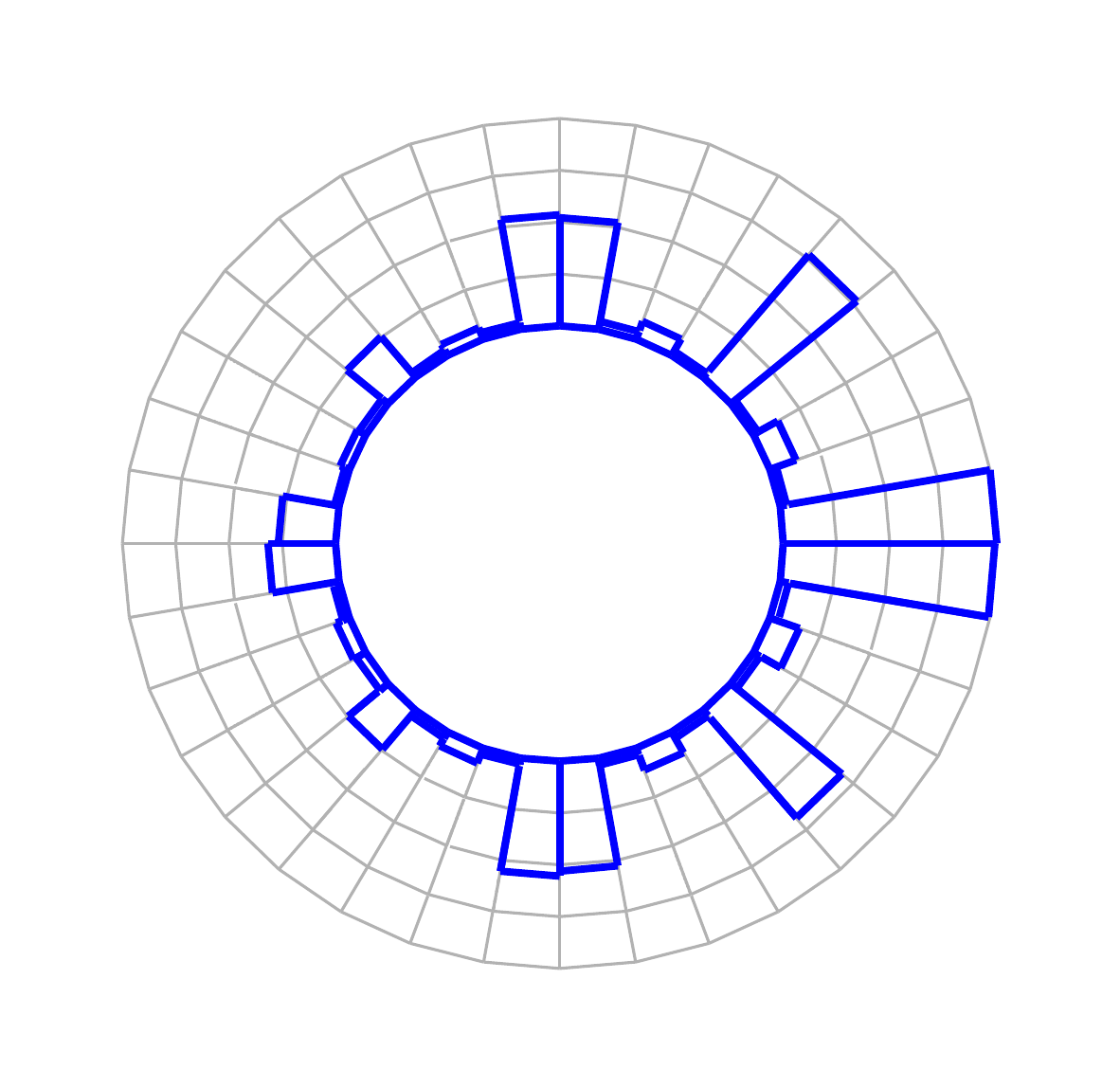}
        \put(91,47){\color{black}0}
        \put(46,90){\color{black}$\pi/2$}
        \put(5,47){\color{black}$\pi$}
        \put(45,5){\color{black}$3\pi/2$}
        \put(38,47){\color{black}Segmented}
    \end{overpic}
    \begin{overpic}[width=0.32\linewidth]{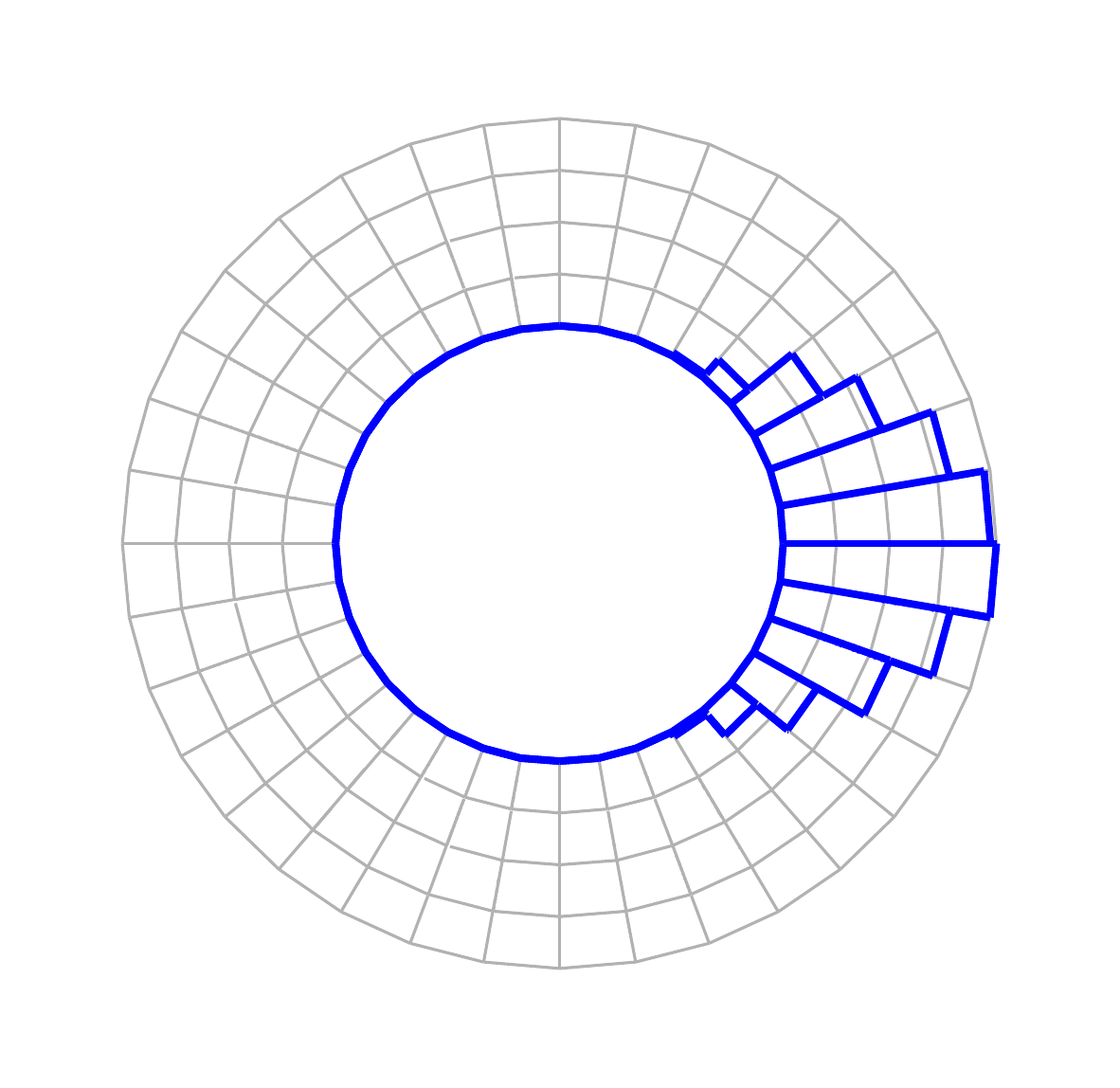}
        \put(91,47){\color{black}0}
        \put(46,90){\color{black}$\pi/2$}
        \put(5,47){\color{black}$\pi$}
        \put(45,5){\color{black}$3\pi/2$}
        \put(38,47){\color{black}Separation}
    \end{overpic}
    \caption{Distributions of the azimuthal angle $\varphi$ for three selected geometries: monolithic (Chooz), segemented (NuLat $5^3$), and separation (Santa). The bin width is 10$^\circ$; the inner circle on each histogram corresponds to 0 events and the outer circle is normalized to the max number of events per bin. }%
    \label{fig:radar}
\end{figure*}
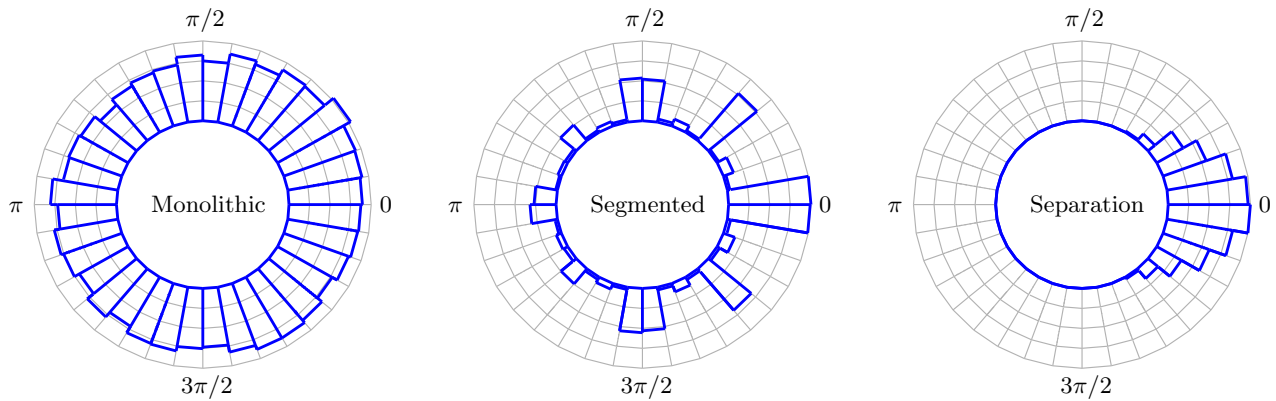

We have chosen a trio of examples which highlight the behaviors of the three basic detector categories, shown in Fig.~\ref{fig:radar}.
The \textit{monolithic}, \textit{segmented}, and \textit{separation} design types are represented by \dc, \nl\-$5^3$, and \santa, respectively.
It is worth noting that, although SANTA is appearing to have superior angular reconstruction, the separation concept in itself is rather a non-realistic approach as it faces major background issues and the packing factor is rather small compared to all other designs --- the target layer is sub percent of the total volume needed for this concept to work.
This study was posited that the incoming antineutrino direction was perpendicular to the detector plane; thus, creating a bias in reconstructing the direction. We studied other orientations of the detector plane (which showed a degraded performance). We decided not to include them here as that would divert the reader's attention toward a rather unrealistic detector concept at the expense of giving other detector designs a similar amount of detail. 

\subsection{\label{subsec:deltaphi}Accuracy and Angular Resolution}

Our primary results are given in Table~\ref{tab:ncomp_results}.
We give the angular resolutions of these experiments in terms of $[\Delta \varphi]_{1\sigma}$, the $1\sigma$ uncertainty on the reconstructed source direction.
It is the half-aperture of a cone in the 3D experiments and the half-angle of an arc in the 2D experiments.
Following the formulation used by Double Chooz~\cite{langbrandtnerThesis,Caden:2012bm,cadenThesis}, we have calculated $[\Delta \varphi]_{1\sigma}$ as follows:

\begin{equation}
  [\Delta \varphi]_{1\sigma} = \mathrm{arctan} \left( \frac{P/d_n}{\sqrt N} \right),
  \label{eq:deltaphi}
\end{equation}
where $[\Delta \varphi]_{1\sigma}$ is the $1\sigma$ angular uncertainty on the reconstructed direction to the \nb\ source, $P$ is the mean position resolution, $d_n$ is the mean distance between the prompt and delayed events, and $N$ is the number of IBD events used for direction reconstruction.
(Recall from \S\ref{subsec:dirn_meth} that $d_n~\equiv~\|\recm\|$).

\begin{equation}
  P = (\sigma_x + \sigma_y + \sigma_z) / 3
  \label{eq:pos_res}
\end{equation}
$P$ is itself calculated via Eq.~\ref{eq:pos_res}, where $\sigma_x$, $\sigma_y$, and $\sigma_z$  are standard deviations for the displacement vector; in the monolithic case: it is the sigma of the Gaussian fit.
For the 2D detectors,
the calculation is as follows:

\begin{equation}
  P = (\sigma_x + \sigma_y) / 2
  \label{eq:pos_res_2D}
\end{equation}

\subsection{\label{subsec:res_agg}Aggregate Results}

\begin{figure}[ht]
  \begin{center}
    \includegraphics[width=1.0\linewidth]{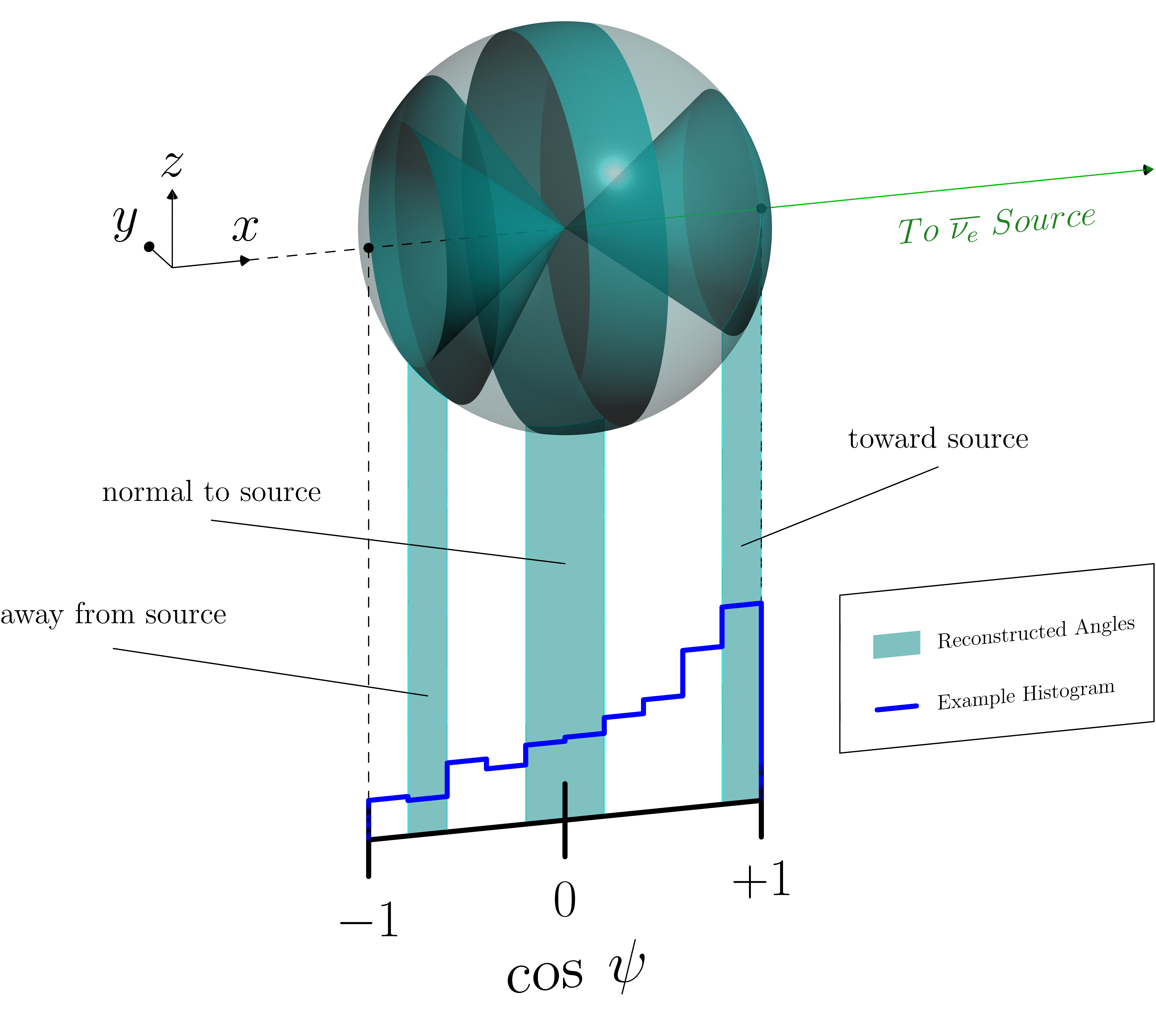}
    \caption{Physical meaning of $\cos\psi$ distributions, illustrated here as it applies to the 3D experiments. The 2D analogue is the projection of this diagram onto the $x$-$y$ plane. Events in the leftmost shaded region (i.e., the second bin from the left in the example histogram) correspond to reconstructed angles falling within the solid angle between the two cones defined by $\cos \psi  = -0.8$ and $\cos \psi  = -0.6$, or $\psi \approx 143^o$ and $\psi \approx 127^o$.}
    \label{fig:cospsi}
  \end{center}
\end{figure}

\begin{figure}[ht]
    \begin{flushright}
    \includegraphics[width=\linewidth]{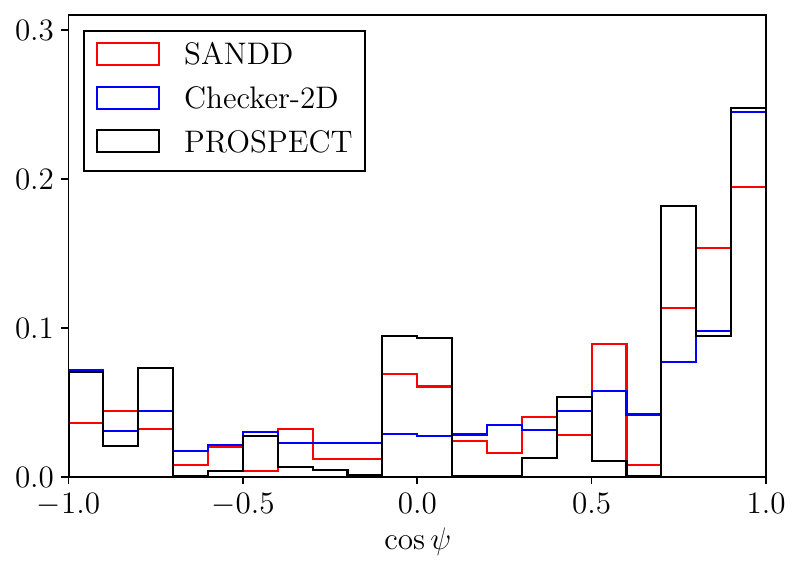}
    \includegraphics[width=\linewidth]{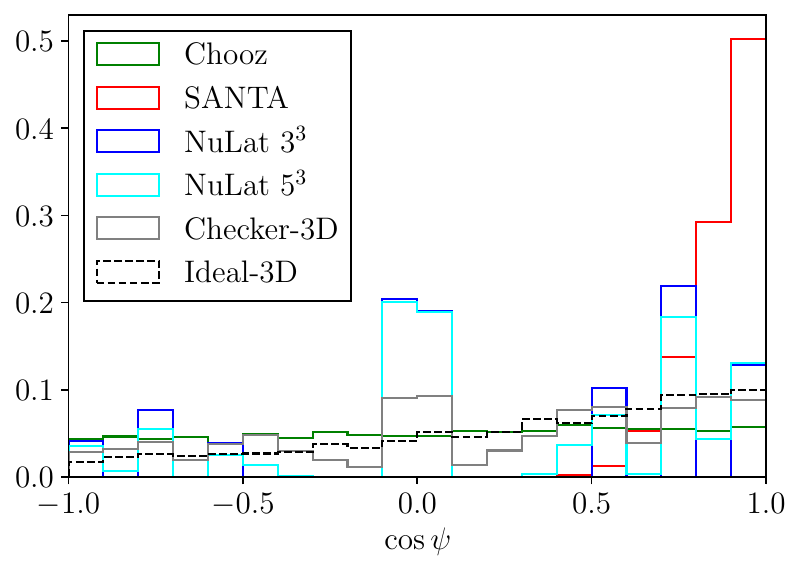}
    \end{flushright}
    \caption{Normalized distribution of $\cos \psi$ for 2D ({\it top panel}) and 3D ({\it bottom panel}) experiments. The Chooz-like monolithic detector is listed among the 3D detectors. The ``Ideal-3D'' distribution indicates the simulated directional performance of a 1.5\%-wt.~$^{6}$Li-loaded, monolithic detector with a hypothetical \emph{perfect} position resolution (\textit{i.e.,} $\delta \{x,y,z\} \equiv 0$). It effectively represents the theoretical maximum directional performance of a non-segmented detector (with this particular target material), due to the neutron-scattering problem discussed in \S\ref{subsubsec:scat}.}
     \label{fig:cospsi_2d_and_3d}
 \end{figure}

The angle $\psi$ \dash or more specifically, its cosine \dash is especially helpful for evaluating a detector's directional performance.
The quantity $\cos \psi $ for a given event gives a value on the interval $[-1,+1]$ that reflects the degree of agreement between \rhat\ and \snu, as illustrated in Fig.~\ref{fig:cospsi}.
Values near $-1$ indicate a reconstructed direction pointing away from the true source direction; values near $0$, perpendicular to the true source direction; and values near $+1$, toward the true source direction.
Stated simply, distributions that are skewed more heavily to the right of these plots generally indicate better directional performance.

\begin{table*}
  \begin{tabular}{ c | l ||  r | r | r || r |r  ||r |r | r || r | r | r }
   Recon. Type & Experiment & Dopant & \%wt. &  Vol.~(L)   & $P$~(mm)  & $d$~(mm) & $ N_{trg}$ & $N_{1V}$ & $N$ &  $\langle \cos \psi \rangle$  & $\varphi~(^\circ)$ &  $\Delta \varphi~(^\circ)$  \\ \hline \hline %

               & DC (data)           &               Gd & 0.1~ &    10~000\textcolor{white}{.00}  & 156.7 & 16.7 &       8249~                  & ---~                  & 8249~                 & 0.059~                &        7.4    & 5.93\\
	       & Monolithic   &               Gd & 0.1~ &   785~000\textcolor{white}{.00}  & 156.7 & 13.9 & 	 9198~			&\textbf{---}~		& 9198~			& 0.052~		&	-2.55    & 6.70\\ 
   \textit{3D} & NuLat $5^3$         &   $\nucl{6}{Li}$ & 1.5~ &       15.6\textcolor{white}{0}   & 43.4 & 20.0 &	 4592~			& 661~			& 3931~			& 0.256~		&	 0.82    & 1.98\\ %
               & 3D Chkbd.           &   $\nucl{6}{Li}$ & 1.5~ &       107.\textcolor{white}{00}  & 105.7 & 32.7 & 	 4384~			& 178~			& 4206~			& 0.194~		&        6.03    & 2.86\\ %
               & SANTA               &                B & 0/5~ &        24.0\textcolor{white}{0}  & 218.28 & 1000.\textcolor{white}{0} &     2275~			& 0~			& 2275~			& 0.876~		&	-0.56    & 0.26\\ \hline %
               & SANDD-CM            &   $\nucl{6}{Li}$ & 1.5~ &         0.64                     & 16.8 & 10.0 &	 250~			& 3~			& 247~			& 0.360~		&	12.08    & 6.11\\ %
   \textit{2D} & 2D Chkbd.           &   $\nucl{6}{Li}$ & 1.5~ &         5.12                     &  55.6 & 24.8 &	 1864~			& 4~			& 1860~			& 0.297~		&	 4.02    & 2.97\\ %
               & PROSPECT            &   $\nucl{6}{Li}$ & 0.08 &     3~465\textcolor{white}{.00}  & 225.7 & 99.0 &	 2257~			& 239~			& 2018~			& 0.335~		&	 3.03    & 2.90 %
  \end{tabular}
  \caption{Summary table. %
    The total number of IBDs simulated in each detector is 10~000. Target material is PVT for all detectors except for DoubleChooz.
  $N_{trg}$ is the total number of IBD-candidate events registered by the antineutrino trigger.
  $N_{1V}$ is the number of IBD-candidate events for which the prompt and delayed events occurred within a single volume/segment; these events were not used for directional reconstruction except in the monolithic (Chooz-type) detector.
  $N = N_{trg} - N_{1V}$ is the resulting number of IBD-candidate events used for directional reconstruction.
  $\langle \cos \psi  \rangle$ is somewhat a measure of angular accuracy, and 
  $\Delta \varphi$ is the $1\sigma$ angular uncertainty on the reconstructed direction to the antineutrino source \dash see \S\ref{subsec:deltaphi}; it is the half-aperture of a cone in the 3D experiments and the half-angle of an arc in the 2D experiments. Note that the 2D experiments are reconstructing the \emph{azimuthal angle only}.}
  \label{tab:ncomp_results}
\end{table*}

A summary of the main results from our primary simulation set is given in Table~\ref{tab:ncomp_results}.
The $\cos \psi $ distributions for the 2D and 3D experiments are shown in Fig.~\ref{fig:cospsi_2d_and_3d}.
The base material (prior to doping) for all, but the monolithic (Chooz-type), detectors is our simulated version of PVT, which is the base polymer for Eljen EJ-254~\cite{Eljen,B10_scint}.
The base material for the monolithic detector is our simulated version of Double Chooz' organic liquid scintillator. %
The column ``Vol'' in the table is the detector's active target volume in liters.  Recall that our Chooz-type tank is substantially larger than those used in the actual \dc\ experiment.  SANTA's capture plane and the inert checkerboard cells are, by definition, not included.
The $1\sigma$ angular uncertainty $[\Delta \varphi]_{1\sigma}$ on the reconstructed direction to the antineutrino source \dash see \S\ref{subsec:deltaphi}; it is the half-aperture of a cone in the 3D experiments and the half-angle of an arc in the 2D experiments.

\begin{figure}[ht]
    \centering
    \includegraphics[width=0.48\textwidth]{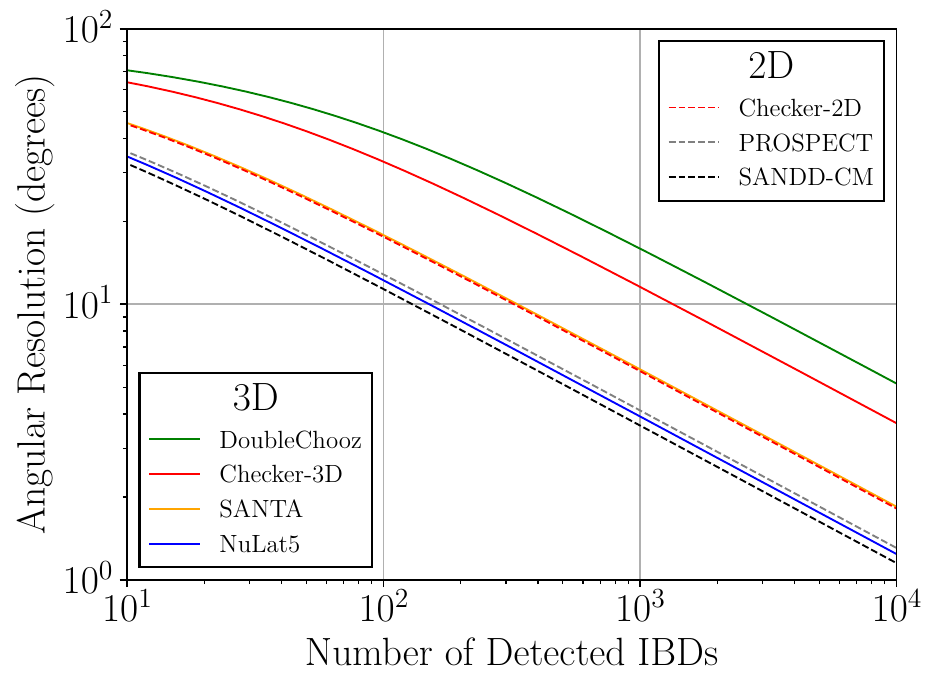}
    \caption{Angular uncertainty $\Delta \varphi$ as a function of detected IBD events for various detector designs (2D detectors indicated by a dashed line; 3D --- by a solid line). The calculation is based on Eq.~\ref{eq_deltaPhi_PackFactor} using the values for the mean position resolution $P$  and the mean distance $d_n$ (between the prompt and delayed events) presented in Table~\ref{tab:ncomp_results} for the simulated detector geometries, as well as taking into account the packaging factor $\Pi$.}
     \label{fig_ang_uncert_PackFactor}
 \end{figure}

To better understand various detector performance and not to give a reader misleading conclusion that SANTA is the best performing detector, we introduce a packaging factor to roughly account for the void/insensitive volume in the detectors. For SANTA, it is only 0.46\%, for Checkerboard-3D and 2D --- 25\% and 50\% respectively, for SANDD-CM --- 70\% while other geometries are assumed to have a packaging factor 100\%. We therefore modify Eq.~\ref{eq:deltaphi} as follows:

\begin{equation} \label{eq_deltaPhi_PackFactor}
  [\Delta \varphi]_{1\sigma} = \mathrm{arctan} \left( \frac{P/d_n}{\sqrt{ N \; \Pi} } \right),
\end{equation}
where $\Pi$ is the packaging factor, and all other definitions kept unchanged.
The resulting angular resolution is represented in Fig.~\ref{fig_ang_uncert_PackFactor}.

\section{\label{sec:concl}Conclusion}

We have shown how a variety of small existing and hypothetical IBD detectors can determine the incoming direction of electron-antineutrinos given a sufficient number of events. Our principal conclusions are as follows:

{

Directionality from near-surface reactor-\ibd\ detectors can be substantially improved with currently-existing technology.
This requires designing the detector geometry to prioritize directional performance.
The same is likely true for \ibd\ detectors in other deployment scenarios as well.
Additionally, bulk-volume \ibd\ detectors \dash\ regardless of segmentation or scale \dash\ are substantially restricted in their potential directional performance, even with continued improvements in vertex resolution.
This is because they are inherently limited by the neutron-scattering problem as detailed in~\S\ref{subsubsec:scat}.
As a result, \ibd\ detectors intended for directional measurements will likely require a geometry incorporating some amount of open (or otherwise neutron-transparent) space for minimally-scattered neutrons to traverse before undergoing thermalization and capture.

The scalability of designs incorporating this effect is limited by two closely-related constraints:
\begin{enumerate}
  \item Along the direction of antineutrino travel, the thickness $d_{active}$ of the detector's target segments should not significantly exceed the neutron-capture distance.
  \item Along this direction, some or all of these active segments should be followed by a sufficiently neutron-transparent region with a thickness $d_{inert}$ that is at least comparable to the neutron-capture distance.
\end{enumerate}
\begin{equation}
    d_{active}~\lesssim~d_n~\lesssim~d_{inert}
    \label{eq:constr}
\end{equation}
These constraints, summarized in Eq.~\ref{eq:constr}, should make it possible for the detector to surpass the performance limits imposed by the aforementioned neutron-scattering problem, with a performance improvement commensurate with the degree to which the constraints are satisfied or exceeded.

3D directional reconstruction is possible but likely not worth the added cost and complexity.
2D reconstruction is much easier to attain and is sufficient for all likely nonproliferation applications of which we are aware.
Nonproliferation scenarios further than $\mathcal{O}(10^2~\mathrm{m})$ from a reactor \dash which are expected to require a~fiducial~mass~$\gg~1~\mathrm{ton}$ \dash will likely be restricted to only 2D reconstruction, largely because of the scalability constraints discussed above.

Checkerboarding is one possible method for hybridizing \santa\ with more robust detectors, but other methods may provide equal or better directional benefit without the added complexity caused by the anisotropy inherent to the checkerboard lattice.
This report is a first exercise in understanding directionality in  different detector designs; a careful background implementation and specific deployment scenarios need to be considered. This will be covered in the follow-on studies.

}

Finally, we would like to conclude with some brief observations on reactor-antineutrino studies in general.
In the near-field case ($5-20$~m standoff, $1-5$~T target mass), a directionally-sensitive detector could study the source distribution of antineutrinos being produced inside the reactor, essentially imaging the reactor core through its shielding and containment structures. 
In the far-field case ($10-100$~km standoff, $1+$~kT target mass), a single detector with directional capability could locate an unknown reactor, a feat which otherwise requires multiple detectors working in concert.
In fact, by combining direction-finding and range-finding, we have seen that at least in principle, one may achieve blind recognition of a reactor's existence, azimuth, distance, and power~\cite{Learned_2008}. %

This study strongly suggests that the field of antineutrino source-finding via \ibd\ detection can make substantial improvements by employing novel geometries that are crafted to maximize directional performance, and we look for the next generation of detectors to expand what we can learn from the only class of elementary particle which preserves information about its point of origin.
On the technology side, although considered challenging in the past,  new techniques have become available to support constructing finely segmented detectors, or detectors of mixed materials, such as 3D printing of $^6$Li or $^{10}$B PSD scintillators~\cite{KIM2023168537}.

\begin{acknowledgments}
This work is supported by the U.S. Department of Energy National Nuclear Security Administration and Lawrence Livermore National Laboratory [Contract No. DE-AC52-07NA27344, release number LLNL-JRNL-859999-DRAFT].
We wish to acknowledge the University of Hawai'i at M\={a}noa, the Department of Physics~\&~Astronomy, and the HEP group in particular.
We also thank engineering consultants Angela~Mareschal and Ed~Reedy.
We thank Gabrielle~Zacek and~Viktor Zacek, the pioneers of the IBD directionality, for sharing the story of how the forward/backward asymmetry in IBD neutrons was first discovered in the G\"osgen experiment.
Last but not least, we acknowledge Adam Bernstein's guidance during the initial stages of this project.
\end{acknowledgments}

\bibliography{Refs}
\bibliographystyle{aipnum4-2} %

\end{document}